\documentclass[aps,prx,groupedaddress,twocolumn,notitlepage,10pt]{revtex4-1}

\usepackage[T1]{fontenc}
\usepackage{array}
\renewcommand{\arraystretch}{1.2}
\usepackage{makecell}

\usepackage{float}
\usepackage{graphicx}  
\usepackage{xcolor}
\usepackage{bm}        
\usepackage{braket}
 \usepackage{amssymb}   
\usepackage{epstopdf}
\usepackage{xfrac}
\usepackage{cancel}
\usepackage{soul}
\usepackage{amsmath}
\usepackage{amsthm}
\usepackage{amsfonts}
\usepackage{bbm}
\definecolor{darkblue}{rgb}{0.1,0.2,0.6}
\definecolor{darkred}{rgb}{0.8,0.1,0.2}
\definecolor{darkgreen}{rgb}{0.31,0.62,0.24}
\usepackage[colorlinks,citecolor=darkblue,linkcolor=darkblue,urlcolor=darkblue]{hyperref} 
\usepackage{enumerate}
\usepackage{url}  
\usepackage{mathrsfs}
\usepackage{mathtools}

\usepackage{multirow}
\usepackage{hhline}

\newcommand{\vac}{\ket{\text{vac}}}
\newcommand{\cnot}[1]{\textsc{cnot}_{#1}}
\newcommand{\cz}[1]{\textsc{cz}_{#1}}
\newcommand{\norm}[1]{\left\lVert#1\right\rVert}

\usepackage[ruled, vlined, linesnumbered]{algorithm2e}

\newenvironment{algo}[1]{
  \algorithm[ht]
    \caption{#1}
    \DontPrintSemicolon
    \SetAlgoCaptionLayout{left}
    \SetAlgoHangIndent{0pt}
    \SetKwInOut{Input}{input}
    \SetKwInOut{Output}{output} 
}{
  \endalgorithm
}

\begin{document}

\title{Quantum Data Center Infrastructures: A Scalable Architectural Design Perspective}

\author{Hassan Shapourian}
\author{Eneet Kaur}
\author{Troy Sewell}
\author{Jiapeng Zhao}
\author{Michael Kilzer}
\author{Ramana Kompella}
\author{Reza Nejabati}
\affiliation{Cisco Quantum Labs, Santa Monica, CA 90404, USA}

\begin{abstract}

This paper presents the design of scalable quantum networks that utilize optical switches to interconnect multiple quantum processors, facilitating large-scale quantum computing. By leveraging these novel architectures, we aim to address the limitations of current quantum processors and explore the potential of quantum data centers. We provide an in-depth analysis of these architectures through the development of simulation tools and performance metrics, offering a detailed comparison of their advantages and trade-offs. We hope this work serves as a foundation for the development of efficient and resilient quantum networks, designed to meet the evolving demands of future quantum computing applications.

\end{abstract}

\maketitle


\section{Introduction}

Quantum computing promises to solve complex problems beyond the reach of classical systems, but realizing its full potential requires the ability to operate millions of qubits. Current quantum processing units (QPUs) are limited to only tens or hundreds of qubits, well below the scale necessary for achieving practical quantum advantage. To bridge this gap, the concept of a quantum data center has been proposed, where multiple QPUs are networked together, enabling a distributed architecture that can scale to meet the demands of large-scale quantum computing~\cite{cacciapuoti2019quantum,caleffi2022distributed,barral2024review}. Ultimately, these quantum data centers will form the backbone of a global quantum network, or quantum internet, facilitating seamless interconnectivity on a planetary scale~\cite{Van_Meter_2022,wehner2018quantum}.

Quantum data centers  (QDCs)~\cite{Liu_2024} present a compelling solution to the limitations of individual quantum processors, leveraging interconnected QPUs to form a distributed quantum computing infrastructure. This model offers not only the scalability needed for large-scale quantum computation but also the economic and operational benefits of centralized quantum resources in a controlled environment. However, the architecture of such a quantum data center must address a variety of challenges, including qubit transfer with a reasonable rate and fidelity, network latency, and the probabilistic nature of quantum entanglement generation and distribution~\cite{sunami2024scalable}.

In this work, we propose scalable architectures for quantum data center networks, inspired by principles of classical data center networking. Our design leverages a dynamic, circuit-switched quantum network to facilitate efficient entanglement distribution between QPUs using shared resources, such as Bell-state measurement devices, quantum memories, and entanglement sources. This approach enables on-demand, all-to-all connectivity across the quantum network while minimizing reliance on expensive quantum hardware, thus optimizing cost and scalability.

Our proposed architectures stand in contrast to one-to-one (peer-to-peer) quantum network designs~\cite{caleffi2022distributed,cuomo2023architectures,He_2024}, where QPUs are sparsely connected via optical links, typically forming a nearest-neighbor topology. As depicted in Fig.~\ref{fig:arch}, we propose connecting QPUs through a non-blocking photonic interconnect composed of optical switches and quantum devices, building on scalable concepts akin to Refs.~\cite{monroe2014large,earl2022architecture,gauthier2023control} and expanded upon in Ref.~\cite{barral2024review}. We explore two categories of quantum network topologies based on classical data center networking paradigms: switch-centric and server-centric. In switch-centric topologies, the network provides direct optical links between every pair of QPUs, achieving full connectivity. Conversely, server-centric topologies offer a modular design with many optical links but without full all-to-all connectivity, positioning them between traditional one-to-one architectures and switch-centric designs.

While one-to-one topologies may suffice for smaller systems—where not every pair of QPUs requires direct physical connections—they become less practical as system size scales to tens or hundreds of QPUs distributed across multiple nodes. For such larger networks, more structured architectures are necessary to maintain high end-to-end fidelity and efficient entanglement generation rates. The modular hierarchy of our proposed architectures enhances scalability and interoperability by accommodating diverse device features and transducers. Additionally, by leveraging dedicated hardware for entanglement distribution, our approach reduces system overhead and unlocks further performance improvements.

We propose several QDC network topologies inspired by classical data center architectures, including Clos, Fat-tree~\cite{6312192}, HyperX~\cite{6375529}, Bcube~\cite{guo2009bcube}, and Dcell~\cite{6770468}, serving as representative examples of switch-centric and server-centric designs. To support entanglement generation, we explore three distinct protocols, enabling QPUs to communicate using communication qubits equipped with spin-photon interfaces~\cite{beukers2024remote}. These interfaces can operate in different modes—emitter, scatterer, or a combination of both—depending on the protocol's requirements. 

To enable efficient execution of distributed quantum computing jobs, we introduce the concept of a \emph{network-aware} quantum orchestrator, a framework designed to bridge physical-layer architectures with higher-level quantum applications in QDC networks. The orchestrator takes circuit-level descriptions of quantum jobs and network topology as inputs and generates precompiled instructions for optical switches and quantum hardware components, which are executed by a classical central controller. 


Additionally, we include a simulation and benchmarking section focusing on circuit execution capabilities~\cite{proctor2024benchmarkingquantumcomputers}, where we evaluate the performance of our proposed architectures using a Clos topology for random quantum circuits~\cite{Knill_2008,Helsen_2022,helsen2021benchmarking} as well as algorithmic benchmarks~\cite{lubinski2023application}. By integrating physical-layer modeling, network protocols, and the network-aware quantum orchestrator, we analyze average network latency and quantum fidelity as key metrics, providing insights into the practical feasibility of our approach.

The rest of our paper is organized as follows: Section~\ref{sec:modeling physical layer} outlines the physical layer modeling approach and details the entanglement generation protocols. In Sec.~\ref{sec:network architecture}, we introduce the proposed QDC network architectures, followed by the presentation of the network-aware quantum orchestrator in Sec.~\ref{sec:orchestrator}. Section~\ref{sec:performance analysis} focuses on numerical simulations and performance benchmarks to evaluate the proposed designs. Finally, Section~\ref{sec:discussion} concludes the paper with closing remarks and potential future research directions. Additionally, four appendices are included to provide extended discussions on QDC architectures, further details on physical layer modeling and simulations, and integer linear programming (ILP) formulations for specific steps of the quantum orchestrator.

\section{Modeling the physical layer}
\label{sec:modeling physical layer}

\begin{figure}
    \centering
    \includegraphics[scale=0.5]{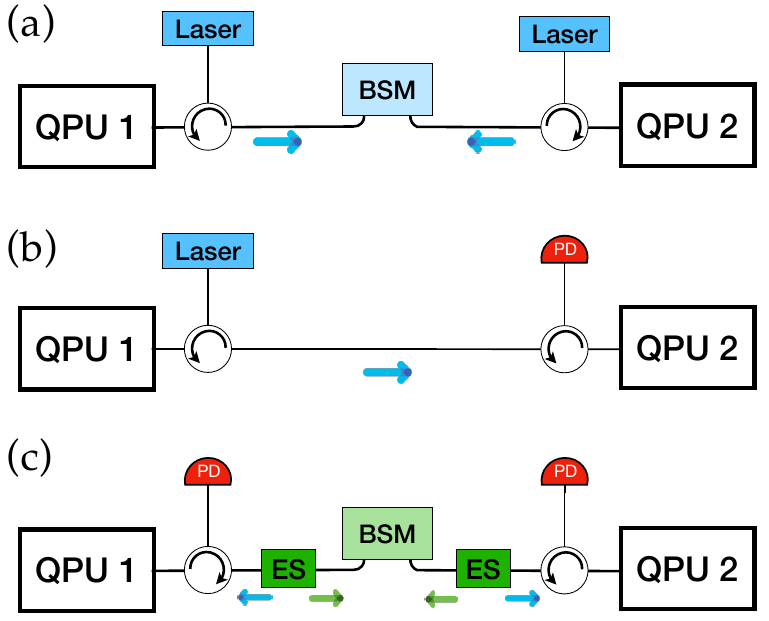}
    \caption{Ebit generation protocols. (a) Emitter-emitter, (b) emitter-scatterer, and (c) scatterer-scatterer protocols. Without any frequency conversion, the (a) and (b) protocols are designed for intra-rack communication, while (c) can be adapted for inter-rack communications. In the latter case, non-degenerate entanglement sources can be used to generate two photons one at telecom and one at frequency ranges compatible with QPUs. }
    \label{fig:ebit-protocols}
\end{figure}

We envision each QPU is equipped with two types of qubits: data qubits, used to carry out quantum computation, and communication qubits, used to generate and store entangled qubits (ebits) between different QPUs.
The goal of QDC network is to enable entanglement generation between communication qubits of various QPUs. The generated entangled pairs are then consumed to execute remote gates between the data qubits of different QPUs. The inter-QPU entanglement pairs can be generated in various ways. We summarize the three most popular methods in Fig.~\ref{fig:ebit-protocols}, which we explain in detail later in Sec.~\ref{sec:Entanglement generation process}. A common theme in these methods is that quantum communication between QPUs is performed by transmitting or receiving photonic qubits in the form of single photon states (in some enconding). The emission process or the scattering process is carried out such that it results in an entangling gate between the stationary (communication) qubit and the flying qubit. This requires an efficient spin-photon interface (see for example the recent review~\cite{beukers2024remote} on various technologies and simulation  tools).
In this section, we start by briefly reviewing the emission and scattering processes and finish by discussing the entanglement generation protocols.

In this paper, we explore on-demand scheduling protocols~\cite{laurat2018demand,chakraborty2019distributed} where ebits are generated dynamically as needed and consumed immediately. This approach allows communication qubits to have lower quality (e.g., shorter coherence times) compared to data qubits, as they only facilitate the generation and immediate consumption of end-to-end entanglement without the need to store quantum information. However, this characteristic does not always hold true. For example, in continuous ebit protocols~\cite{inesta2023performance,talsma2024continuously}, where a buffer of EPR pairs is maintained and ebits are consumed based on program requirements, high-quality communication qubits (or reliable quantum memories) are essential to preserve the fidelity of the end-to-end ebits over time.

Before getting into details of how entanglement is generated between flying photonic qubits and communication qubits, let us briefly review different encodings and present our notation. We denote the vacuum state by $\vac$ and the photonic mode creation and annihilation operators by $a$ and $a^\dag$, respectively.
\begin{itemize}
    \item Fock-space: 
    This is also called presence/absence encoding. We identify the two states of the photonic qubit as
    \begin{align}
        \label{eq:fock-bin}
        \ket{1}_p\coloneqq a^\dag \vac, \qquad \ket{0}_p\coloneqq \vac.   
    \end{align}
    \item Polarization: 
    The computational basis for this encoding is defined as 
    \begin{align}
        \label{eq:pol-bin}
        \ket{1}_p\coloneqq a_v^\dag \vac, \qquad \ket{0}_p\coloneqq a^\dag_h \vac, 
    \end{align}
    where the subscript $h(v)$ refers to a horizontally (vertically) polarized photon, respectively.
    \item Time-bin: 
    The computational basis for a time-bin photonic qubit is defined as 
    \begin{align}
        \label{eq:time-bin}
        \ket{1}_p\coloneqq a_e^\dag \vac, \qquad \ket{0}_p\coloneqq a^\dag_l \vac,    
    \end{align}
    where the subscript of the photon creation operator $e(l)$ denotes the early (late) generation time, respectively.
\end{itemize}

We note that the polarization and time-bin encodings can be converted into each other by using linear-optic components such as polarizing beam splitter and fiber delay lines. As we explain below, ebit generation process ends with measuring photonic qubits using single-photon detectors, which is in turn used as a heralding signal for a successful attempt. 
However, different measurement modules are required for different encodings, and sometimes it is not possible to measure qubits in an arbitrary basis. For instance, the Fock space encoding
only allows for measurement in the computational basis ($\ket{0}$, $\ket{1}$) due to superselection rules (i.e., there cannot be a basis to measure a qubit in a superposition of one photon and zero photon).


\subsection{Communication qubit-photon interface}

\begin{figure}
    \centering
    \includegraphics[scale=0.5]{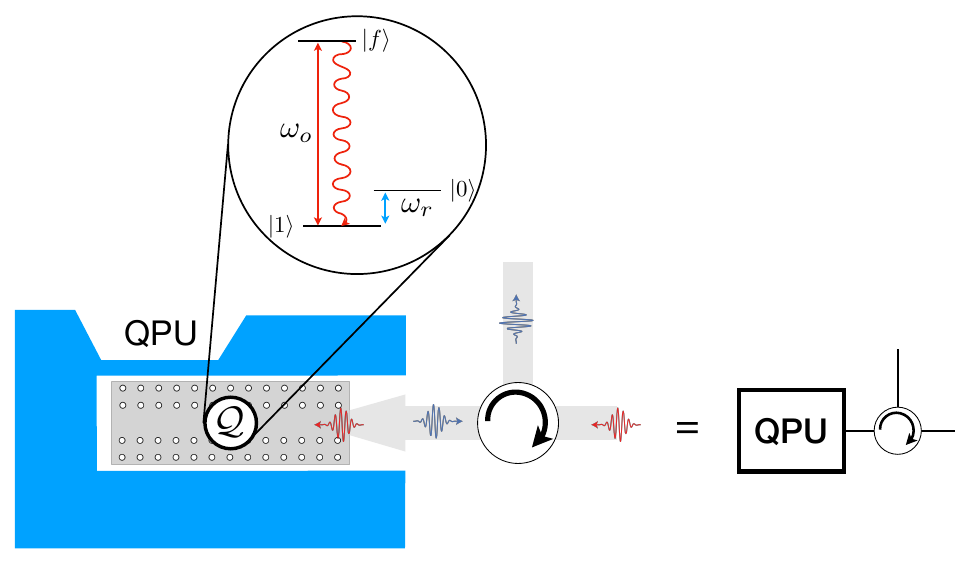}
    \caption{An abstraction for the quantum communication ports of a QPU based on the interface between the communication qubit and photonic qubit. The optical circulator is used to separate the incoming (red) photonic qubit or coherent pulse from the (blue) outgoing photonic qubit. The energy level structure of communication qubit is shown above.}
    \label{fig:comm-qubit}
\end{figure}

We assume the communication qubit is characterized by $\Lambda$-type energy levels and only the transition $\ket{1} \leftrightarrow \ket{f}$ is active as shown in Fig.~\ref{fig:comm-qubit}. Majority of our discussion and protocols can be easily adapted to other types of the energy levels such as the II-type (where there are two allowed optical transitions one for each logical state to transition to). Throughout this paper, we may use spin or stationary qubit to refer to the communication qubits and use photon or flying qubit to refer to the photonic qubit. 
We run the communication qubit in two modes: as a scattering center for incoming photons, and as a quantum emitter. 

In what follows, we use the subscripts $c$ and $p$ to refer to the communication qubit and the photonic qubit, respectively.

\subsubsection{Communication qubit as a quantum emitter}

The high-level idea here is that by manipulating the initial state of the communication qubit we can entangle the outgoing photon to the emitter via the emission process.
The protocol here follows one step of the celebrated Linder-Rudolph protocol~\cite{lindner2009proposal} and is a common approach in trapped-ion qubits~\cite{monroe2014large} and superconducting quits~\cite{ferreira2024deterministic}.

In general, we initialize the communication qubit in a superposition state $\ket{\alpha}_c =  \sqrt{\alpha} \ket{1}_c + \sqrt{1-\alpha} \ket{0}_c$ by sending a resonant pulse between $\ket{0}\leftrightarrow \ket{1}$ at $\omega_r$ where the pulse duration determines the superposition coefficient.
Then, we send a resonant $\pi$-pulse of $\omega_o$ which after possible spontaneous emission gives the following state
\begin{align}
\label{eq:bell-fock}
\sqrt{\alpha}  \ket{1}_c \otimes a^\dag \vac + \sqrt{1-\alpha} \ket{0}_c \otimes \vac.
\end{align}
Given the Fock space computational basis in Eq.~(\ref{eq:fock-bin}) for the photonic qubit, Eq.~(\ref{eq:bell-fock}) describes an entangled state of photonic and communication qubits. 

Similarly, to create an entangled state with a time-bin flying qubit, we initialize the emitter state in a Hadamard state $\ket{+}_c$ (corresponding to $\alpha=1/2$) by sending a resonant $\pi/2$-pulse at $\omega_r$.
Then, we send a resonant $\pi$-pulse of $\omega_o$ which after emission yields the state
\begin{align}
    \label{eq:bell-early}
 \frac{1}{\sqrt{2}} ( \ket{1}_c \otimes a_e^\dag \vac + \ket{0}_c \otimes \vac).
\end{align}
Note that the photon emitted at this instance corresponds to the early time bin.
Next, we apply a $\pi$-pulse of $\omega_r$ followed by another  $\pi$-pulse of $\omega_o$ where the state after the emission is found to be
\begin{align}
 & \frac{1}{\sqrt{2}} ( \ket{0}_c \otimes a_e^\dag \vac + \ket{1}_c\otimes  a_l^\dag \vac) \nonumber \\
&= \frac{1}{\sqrt{2}} ( \ket{0}_c \otimes \ket{1}_p + \ket{1}_c \otimes \ket{0}_p),
\end{align}
which is a Bell pair between the emitter and the photonic qubit. After another $\pi$-pulse of $\omega_r$ we can convert it to another form
\begin{align}
\label{eq:bell-time-bin}
 \frac{1}{\sqrt{2}} ( \ket{1}_c\otimes \ket{1}_p + \ket{0}_c \otimes \ket{0}_p),
\end{align}
if desired.
It is worth noting that a spin-photon entangled state with a polarization encoding can be generated using a communication qubit with II-type energy levels where each transition is coupled to a particular polarization. In this case, the resonant $\pi$-pulse is prepared in a superposition of the two polarizations~\cite{zhan2020deterministic}.

To sum up, we initialize the state of communication qubits and arrange signals to make
the emission process effectively a $\cnot{c,p}$ gate where the control qubit is the emitter~\cite{shapourian2023modular}.

 \subsubsection{Communication qubit as a scattering center}
\label{sec:Communication qubit as a scattering center}

As we see in Fig.~\ref{fig:ebit-protocols}, the (b) and (c) methods involve receiving a photonic qubit which after going through the QPU (which we call scattering process) is detected at photon detector (PD) modules. The scattering process is engineered to let the incoming photon and the communication qubit interact and effectively realize an entangling gate (aka, spin-photon gate). At the same time, the detection event need to perform a measurement in a superposition basis, e.g., Hadamard basis, to ensure transferring the entanglement to the scatterer. Because of that, the entanglement generation protocols based on scattering events are not applicable to the Fock-space encoding, where photonic qubits can only be measured in the computational basis.
In what follows, we briefly discuss such effective gates.

The first approach implies a deterministic controlled-phase ($\cz{}$) gate~\cite{pichler2017universal,zhan2020deterministic} and works as follows: If the communication qubit is at $\ket{1}$, an incoming photon at frequency $\omega_o$ is absorbed and reemitted back. The reemission process involves a full rotation in the two-dimensional subspace ($\ket{1}$-$\ket{f}$) and gives a $\pi$ Berry phase. 
Because of this phase shift, a scattering event can be described by the following map
\begin{align}
    \ket{0}_c\otimes  \vac &\to \ket{0}_c\otimes \vac, \nonumber \\
    \ket{0}_c\otimes a^\dag \vac &\to \ket{0}_c\otimes a^\dag \vac, \nonumber \\
    \ket{1}_c\otimes  \vac &\to \ket{1}_c\otimes  \vac, \nonumber \\
    \ket{1}_c\otimes a^\dag \vac &\to -\ket{1}_c\otimes a^\dag \vac, 
\end{align}
which is nothing but a $\cz{}$ gate considering the encoding defined in Eq.~(\ref{eq:fock-bin}).
Such spin-photon gates are demonstrated in atoms trapped in Fabry-Perot cavities~\cite{daiss2021quantum}, 
and quantum dots in photonic crystal waveguides~\cite{lodahl2017chiral,chan2022quantum}.
Similar to the emission process, this method can be readily adapted to time-bin encoding. Suppose we want to create a Bell state of the photonic and communication qubits.
The incoming photonic qubit and the communication qubit are both initialized in the superposition state
$\ket{+}_c \otimes \ket{+}_p = \frac{1}{2} (\ket{0}_c + \ket{1})\otimes (a_e^\dag + a_l^\dag)\vac$. 
Since photon scattering always give a $\pi$-phase shift, a simple way of implementing the controlled-phase gate is by only letting the early time-bin interact with the communication qubit via an 1-to-2 optical switch (e.g., a phase-tunable Mach-Zehnder interferometer) and rerouting the late time-bin. This leads to the following Bell state
\begin{align}
    \frac{1}{\sqrt{2}} (\ket{+}_c\otimes \ket{0}_p+\ket{-}_c \otimes\ket{1}_p ).
\end{align}
A drawback of the scattering process is that approach requires a strong coupling of photons with the communication qubit. Although strong coupling regime can be realized by confining the light with photonic cavities or waveguides, this technology may not be available in all platforms.

Another approach is based on the conditional photon reflection, which is also referred to as ``carving''.
This process is based on the fact that a photon can only be reflected if the communication qubit is in $\ket{1}$ state (c.f.~Fig.~\ref{fig:comm-qubit}), and hence a successful gate is heralded by detection of a reflected photon. This approach is experimentally realized in SiV centers in diamond photonic crystal cavities ~\cite{nguyen2019quantum,bhaskar2020experimental,knaut2024entanglement}
and entangling two neutral atoms in a cavity~\cite{welte2017cavity}.
To illustrate how this approach works, we again initialize the system with the product state $\ket{+}_c \otimes \ket{+}_p = \frac{1}{2} (\ket{0}_c + \ket{1}_c)\otimes (a_e^\dag + a_l^\dag)\vac$. 
Right after the early time-bin passes, we apply a $\pi$-pulse of $\omega_r$ to flip $\ket{0}_c$ and $\ket{1}$. A detection event then projects the state into an entangled state $\frac{1}{\sqrt{2}} (\ket{0}_c \otimes a_e^\dag \vac + \ket{1}_c \otimes a_l^\dag \vac)$. This is because only state $\ket{1}_c$ can reflect the photon. Compared to the deterministic spin-photon gates, Bell carving is simpler to realize as it does not require strong coupling. However, this process is inherently probabilistic as it involve post-selection. In other words, the maximum success probability of the Bell carving is 50\% even in the absence of other causes for the photon loss.


\subsection{Entanglement generation protocols}
\label{sec:Entanglement generation process}

In this part, we explain the three main entanglement generation protocols as outlined in Fig.~\ref{fig:ebit-protocols}. We shall refer to these protocols using the operating modes of the two communication qubits in the two end-QPUs as follows: Emitter-emitter, emitter-scatterer, and scatterer-scatterer. We further discuss the performance of each protocol in terms of the end-to-end entanglement generation rate and fidelity following the earlier work~\cite{jiang2007distributed,monroe2014large,ang2024arquin} across different platforms.

In short, the emitter-emitter protocol involves both communication qubits running as emitters and the photons being directed to a Bell-state swapping module (BSM) to perform entanglement swapping (and a heralding signal). In the emitter-scatterer protocol, 
one communication qubit runs as an emitter and the other as a scatterer, where the heralding signal is generated from a photon detector connected to the second QPU.
 In the scatterer-scatterer protocol, both communication qubits run as scatterers and non-degenerate entanglement sources and BSMs are utilized to distribute entanglement. We further discuss which hardware platforms are more suitable for which protocols. 
We recall that the latter two protocols are not applicable to the Fock-space enconding of photonic qubits as discussed in Sec.~\ref{sec:Communication qubit as a scattering center}.


\subsubsection{Emitter-emitter protocol}
\label{sec:emitter-emitter}

This protocol is shown in Fig.~\ref{fig:ebit-protocols}(a), where we drive both communication qubits in the two end QPUs to emit photons and then post-select the two-qubit measurement outcome of the BSM in the middle. The BSM is often realized by linear optics and single photon detectors, and as such their success probability is $50\%$ regardless of the qubit encoding. Boosted BSM with linear optics can surpass the $50\%$ limit but requires additional ancillary photons~\cite{grice2011arbitrarily,ewert20143,bayerbach2023bell}.
Considering the Fock-state encoding of photonic qubits, where the initial communication qubit-photon entangled state is given by Eq.~(\ref{eq:bell-fock}), the BSM can be implemented by a beam splitter with two single-photon detectors at the two output ports. A single photon detection in either detectors heralds the creation of the state
\begin{align}
    \label{eq:cc_epr}
    \ket{\phi_\pm}_{c_1c_2} = \frac{1}{\sqrt{2}} (e^{ik\Delta x} \ket{1_{c_1}0_{c_2}}\pm \ket{0_{c_1}1_{c_2}}),
\end{align}
with the success probability of $2\alpha(1-\alpha)$, where we group together the communication qubit states, $\ket{q_{c_1} q'_{c_2}} = \ket{q}_{c_1}\otimes \ket{q'}_{c_2}$, and $k\Delta x$ arises due to differences in optical path lengths~\cite{monroe2014large}.
We note that the success probability is further reduced due to photon loss during transmission from each QPU to the BSM.
In particular photon loss may cause a false positive signal. As a result, a single photon detection event yields a noisy Bell state in the form of $\hat \rho_\pm = (1-w) \ket{\phi_\pm}\bra{\phi_+}+ w \ket{11}\bra{11}$, and the resulting fidelity is then given by $F^\text{(F)}_\text{ee} = 1-w$, where
\begin{align}
    w = \frac{\alpha (1-\eta)}{1-\alpha \eta}
\end{align}
and the overall success probability is found to be 
\begin{align}
    p_\text{ee}^\text{(F)} = 2 \alpha \eta (1-\alpha \eta),   
\end{align}
with $\alpha$ the initial state parameter defined in Eq.~(\ref{eq:bell-fock}) and $\eta = \eta_\text{eb}\eta_\text{det}$ the overall end-to-end photon transmission rate (i.e., overall photon loss rate is $1-\eta$) including the detection efficiency of single-photon detectors $\eta_\text{det}$ and the transmission probability from the emitter to the BSM $\eta_\text{eb}$. As we see from the above expressions, in the lossy channel regime $\eta\ll 1$ the larger $\alpha$ leads to higher success probability at the cost of lower fidelity. Therefore, depending on the application and system characteristics we may choose values different from $\alpha=1/2$ in Eq.~(\ref{eq:bell-fock}).
In the above analysis, we neglected optical path difference which is usually a good assumption provided that the optical path difference between two emitters are at the wavelength level of photonic qubits; otherwise, the mismatch leads to a phase factor and the imbalance in transmission rate along the two paths  $\eta_{e_1 b}\neq \eta_{e_2 b}$ results in the entanglement between the two emitters being a noisy Bell state, as they are not maximally entangled.



Because of the probabilistic nature of the EPR pair generation, we consider a repeat-until-success protocol, where we keep trying to generate an EPR pair until we get the positive signal in the BSM. Mathematically, the number of trials is a random variable $N$ described by the geometric distribution $P(N=n) = p_\text{ee}^\text{(F)}  (1-p_\text{ee}^\text{(F)} )^{n-1}$ and the duration time is $N \tau_0$, where $\tau_0$ is the operation time for each attempt. Hence, the average time for a successful EPR pair generation is $\overline{N} \tau_0=\tau_0/p_\text{ee}^\text{(F)} $, which implies the average generation rate
\begin{align}
    R_\text{ee}^\text{(F)} = \frac{2 \alpha \eta_\text{eb}\eta_\text{det}}{\tau_0} (1-\alpha \eta_\text{eb}\eta_\text{det}).
\end{align}
We provide some back of envelope estimate of the resulting rate and fidelity in the next section.

It is important to note that the emitted photons in the above protocol run at the qubit resonant frequency which are typically in the visible or near-infrared (NIR) frequency range ($700$-$900$nm) assuming atomic based or trapped ion quantum computing platforms. Therefore, this protocol by construction (unless we perform quantum frequency conversion to telecom range) is only suitable for short-range quantum communication such as the same-rack entanglement pair generation. We comment more on this issue in the next section as we present the network architectures.

As mentioned, the emitter-emitter protocol based on the Fock-space encoding is sensitive to optical path difference. These challenges can be mitigated by adapting the protocol to time-bin encoding. In this case, the initial spin-photon entangled state is given by Eq.~(\ref{eq:bell-time-bin}) and the BSM must split the time bins to path encoding before sending them to two beam splitters (associated with early and late time bins) each with two single-photon detectors at their output ports. Unlike the Fock-space encoding, a coincidence event in two detectors each of which attached to a different beam splitter heralds the creation of the state in the same form as Eq.~(\ref{eq:cc_epr}), not a mixed state. In other words, there is no false positive event with time-bin (or polarization) encoding if we neglect the detector's dark count, and the nominal fidelity of the heralded states can reach unity even in the presence of photon loss.
As mentioned, a successful event requires arrival of two photons (each with probability $\eta_\text{eb} \eta_\text{det}$) and generating a time-bin spin-photon entangled state takes $\tau_0 + \tau_\text{b}$ where $\tau_\text{b}$ denotes the time difference between the two time bins. Similar to the Fock-space encoding the randomness of the generation process is described by a geometric distribution with the success probability $p_\text{ee}^\text{(T)}= \eta_\text{eb}^2\eta_\text{det}^2$.
Therefore, the entanglement generation rate on average is given by
\begin{align}
    R_\text{ee}^\text{(T)} = \frac{\eta_\text{eb}^2\eta_\text{det}^2}{2 (\tau_0+\tau_b)},
\end{align}
where the factor of $2$ in the denominator is due to the post-selection of the measurement outcomes in the BSM. 



\subsubsection{Emitter-scatterer protocol}

This protocol is shown in Fig.~\ref{fig:ebit-protocols}(b), where the first (emitter) communication qubit is coherently driven to emit a photon which is then received by the other (scatterer) communication qubit and ultimately measured in the photon detector. 
As mentioned, this protocol is not applicable to Fock-space encoding of photonic qubits, so we  consider the time-bin encoding for example, where the initial emitter-photon state is given by Eq.~(\ref{eq:bell-time-bin}). A successful heralding event at the scatterer thus leads to a Bell-pair of the two communication qubits as in Eq.~(\ref{eq:cc_epr}). 
 A successful event here requires the detection of the emitted photon which implies $p_\text{es}^\text{(T)} = \eta_\text{eb} \eta_\text{det}$ ($\eta_\text{es}$ denotes the transmission probability from the emitter to the scatterer) and takes spin initialization and state preparation of $\tau_0 + \tau_\text{b}$. Hence, the average end-to-end entanglement generation rate is found to be
\begin{align}
    R_\text{es}^\text{(T)} = \frac{\eta_\text{es}\eta_\text{det}}{2 (\tau_0+\tau_b)},
\end{align}
where the factor of $2$ in the denominator is due to the post-selection of the measurement outcomes using the Bell carving scheme (c.f.~\ref{sec:Communication qubit as a scattering center}). 
We note that the nominal fidelity of the heralded states in this protocol can reach unity since there is no false positive event if we neglect the detector's dark count.
 Since only a single photonic qubit is transmitted through the network, this protocol imposes fewer requirements for qubit stabilization and synchronization compared to the emitter-emitter protocol. Consequently, the emitter-scatterer protocol is more advantageous in scenarios where stabilizing and synchronizing multiple photonic qubits pose significant challenges.
 


\subsubsection{Scatterer-scatterer protocol}
\label{sec:scatterer-ccatterer}


As shown in Fig.~\ref{fig:ebit-protocols}(c) this protocol starts with two entanglement sources generating two entangled pairs of photons and direct one photon to the BSM in the middle and send the other photon to the end-QPUs. A successful attempt of generating an end-to-end entanglement is heralded by the simultaneous occurrence of three detection events: Two scattering detection events and one coincidence event at the BSM.
The use of entanglement sources in this protocol offers both advantages and challenges. One notable benefit is the utility of non-degenerate sources, which generate pairs of entangled photons at distinct frequency ranges—for instance, near-infrared (aligned with the communication qubit's resonant frequency) and telecom (compatible with optical fibers and standard off-the-shelf devices). This pairing facilitates interfacing between two remote QPUs without relying on underdeveloped quantum frequency converters or transducers.

However, the most commonly available entanglement sources, such as those based on spontaneous parametric down-conversion (SPDC) or spontaneous four-wave mixing (SFWM), have an inherently probabilistic generation process. The stochastic nature of photon pair generation, combined with the requirement for three successful detection events, results in significantly low end-to-end ebit generation rates. While synchronization challenges can be mitigated using quantum memories at the BSM stage, our analysis below demonstrates that the advantages of this protocol—even without quantum memories—justify its adoption in the near term.

We consider the entanglement sources generate a time-bin entangled state of photons
\begin{align}
    \ket{\text{ES}_i} = \frac{1}{\sqrt{2}} ( a_{i,t_i}^\dag b_{i,t_i}^\dag  + a_{i,t_i+\tau_b}^\dag  b_{i, t_i+\tau_b}^\dag  ) \vac,
\end{align}
where $a^\dag$ and $b^\dag$ denote the creation operator of the two entangled photons (aka signal and idler photons), and the subscripts contain two parts: the first index $i=1,2$ refers to the output of the first and second entanglement sources, respectively, and $t_i$ is the $i$-th source photon wavepacket's (mean) characteristic time (c.f.~Eq.~(\ref{eq:photon-packet}) in Appendix~\ref{app:scatter-scatter}). Similar to before, $\tau_\text{b}$ denotes the time difference between the two time bins. A possible way to generate time-bin entangled pair of photons is by splitting an input pulse into two pulses (or bins) by sending it through an interferometer before entering the nonlinear medium~\cite{orieux2017semiconductor}. 
The fact that the generation time is stochastic implies that $t_i$ is a random variable. Without making any assumption about details of the entanglement sources, we consider the pair generation to be governed by a Poisson distribution with an average rate $\lambda$, which physically corresponds to the effective end-to-end (source-to-detector) rate. 

Without the use of quantum memories, we propose a brute-force protocol by which we continuously pump the entanglement sources and look for coincident events across the aforementioned three end points. If we observe a detection event at any of the QPUs but not all three locations, then we reinitialize that communication qubit and reject any events during the reinitialization. 
The underlying reason that this approach gives a finite end-to-end ebit generation rate is that the photon wavepackets have some linewidth  $\Delta \omega$ in frequency domain (or broadening in time) which leads to some finite probability for the coincidence as long as $|t_1-t_2|\lesssim \Delta \omega^{-1}$.
To capture this effect accurately, we numerically simulate this protocol and study the statistics of the time takes to observe a successful event (see Appendix~\ref{app:scatter-scatter} for details). We find that the end-to-end generation time $T_\text{ss}$ follows an exponential distribution 
\begin{align}
    \label{eq:exp-dist}
    P(T_{ss}=x)= \lambda_\text{ss} e^{-\lambda_\text{ss} x}     
\end{align}
where the average ebit generation rate is nothing but the exponential distribution parameter $R_{ss}= \frac{1}{\overline{T_\text{ss}}} = \lambda_\text{ss}$. 
We observe that the parameter $\lambda_\text{ss}=f(\tau_0,\Delta\omega)$  generally varies as we change the photon linewidth and the qubit initialization time. Check out Appendix~\ref{app:scatter-scatter} for plots showing these functional dependencies.

In principle, the nominal fidelity of the heralded states in this protocol can also reach unity since there is no false positive event provided that we neglect the detector's dark count.







\section{Network architecture designs}
\label{sec:network architecture}

\begin{figure*}
    \centering
    \includegraphics[scale=.7]{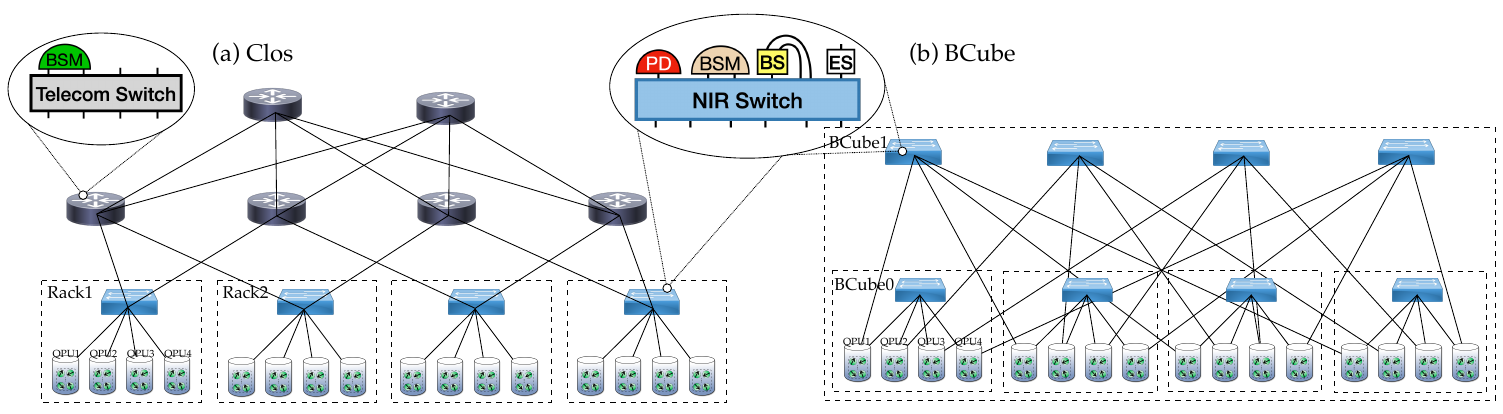}
    \caption{Quantum data center network architectures: (a) Clos topology, (b) BCube topology, as a representative architecture for switch-centric and server-centric topologies, respectively. We use two types of switches: Telecom switches (dark blue disk-shaped) and near-infrared switches (blue rectangular cubes). Lines connecting switches are optical fiber bundles.}
    \label{fig:arch}
\end{figure*}

We envision a network of interconnected QPUs to not only increase the scalability of computing but also facilitate the maintenance of the stringent physical conditions QPUs demand.  
  We base our design on a set of guiding principles to achieve a modular, scalable, and non-blocking quantum interconnect.
As we explain further in this section, our architecture utilizes an optical network fabric to distribute entanglement among QPUs. 
 
First, we look for modular designs with an effective on-deamnd all-to-all connectivity while saving on number of expensive quantum hardware.  
As a result, our architecture involves a dynamic circuit-switched network where the ebits are generated across the quantum network by a limited set of shared resources such as BSMs, entanglement sources, quantum memories, etc. Second, for quantum network topology design, we draw inspiration from classical data center networks~\cite{guo2022} based on switch-centric and server-centric configurations.
Third, we make informed decisions on the type of ebit generation protocol suitable for various topologies and over different length scales.
As we discuss in the next section, we consider a classical control plane in charge of reconfiguring the optical switches and reserving the necessary quantum network devices to create an end-to-end optical paths between the communication qubits and run multiple attempts to generate ebits.

  In what follows, we dive deep into two types of architectures for DQCs as shown in Fig.~\ref{fig:arch}.
In classical data center networks, switch-centric topologies depend on switches for interconnection and routing, whereas server-centric topologies utilize servers equipped with specialized network interface cards (NICs) to facilitate interconnection and routing.
A general observation is that the former mainly scale up in a vertical manner, i.e., increasing the number of switch ports or increasing port speed directly, which leads to a high cost for a large-scale data center. In contrast, the latter 
offer a more flexible network innovation and customization due to the openness and programmability of server hardware and software. We believe that this difference applies to the quantum version of these architectures as well. However, a new challenge which is intrinsically quantum arises in server-centric quantum networks: The interconnection and routing involve a repeater chain for a subset of QPU pairs, where some QPUs play the role of quantum repeaters and require to perform entanglement swapping to generate an end-to-end ebit. We expand more on this below.

\subsection{Switch-centric networks}




Figure~\ref{fig:arch}(a) shows a Clos network as an example of switch-centric topologies. The QPUs are grouped together in several racks which are internally connected via top-of-rack (ToR) switches.
The racks are in turn connected via a hierarchically connected set of optical switches.
The switches are equipped with various quantum hardware devices to enable the ebit generation protocols.
To address the frequency mismatch between the qubit resonant frequencies and telecom wavelengths, we consider short-distance (intra-rack) communications to operate at native qubit frequency (usually at $700$-$900$nm, i.e., near infrared regimes for atomic or ionic platforms) while long-distance (inter-rack) communications to operate at the telecom wavelengths.
In other words, the upper part of the network above the ToR switches (as shown in Fig.~\ref{fig:arch}(a)) operates at the telecom range while the lower part within each rack operates at the NIR regime.

We consider emitter-emitter or emitter-scatterer protocols for the intra-rack ebit generation and scatterer-scatterer for the inter-rack ebit generation. As a result, the ToR switch and all components attached to it run at NIR frequency range, except the entanglement sources the output of which going to the ToR switch runs at NIR and the other output runs at telecom. Therefore, we must use non-degenerate entanglement sources for this purpose. Alternatively, the conversion from NIR to telecom at the ToR switch ports can be done via quantum frequency converters(QFCs)~\cite{van2022entangling,saha2023low,knaut2024entanglement,bersin2024telecom,bersin2024development}. To extend the communication range to other racks, we consider converting the NIR photon into telecom regime and back via QFCs. 

The rationale behind this choice of hybrid operation is that we prefer to maintain a small overhead for moderate quantum jobs, which can be done by a small number of QPUs, since the protocols involving QFCs or non-degenerate entanglement sources may be slow and/or have low fidelity although these technologies are under rapid development. We briefly discuss the quantum adaptation of other popular switch-centric topologies such as Fat-tree and HyperX in Appendix~\ref{app:hyperx} and summarize the number of network components in Table~\ref{tab:network-params}.

\subsection{Server-centric networks}

As Fig.~\ref{fig:arch}(b) shows, a server-centric topology is characterized by smaller switches with fewer ports but QPUs with multiple ports. 
Here, we illustrate a novel architecture inspired by the BCube topology~\cite{guo2009bcube}. We note that server-centric topologies entirely operate at the NIR frequency range, and the ebit generation protocol of choice is the emitter-emitter method although the emitter-scatterer method may be necessary (e.g., for DCell topology). 
As mentioned earlier, the major challenge with such topologies is that they do not provide direct optical paths between every pair of QPUs, e.g., QPU$_i$ and QPU$_j$ of two different BCube containers. For instance, in order to establish entanglement between QPU$_1$ of BCube0 and QPU$_2$ of BCube1, we first need to generate two elementary link entanglements: one between QPU$_1$ of BCube0 and QPU$_1$ of BCube1, and the other between QPU$_1$ and QPU$_2$ of BCube1. This is nothing but the first generation repeater networks~\cite{muralidharan2016optimal}, where intermediate QPUs play the role of quantum repeaters.
In other words, routing and resource management in server-centric topologies involve dealing with repeater networks which is a vast topic by itself (see for example a recent survey~\cite{abane2024entanglement} and references therein). That said, compared to long-distance quantum communication and repeater networks in general, QDC-scale network enjoys an efficient global control plane and does not deal with an arbitrary network graph. Furthermore, the entanglement swapping can be made deterministic because QPUs are in principle capable of applying deterministic gates between their communication qubits. In fact, further simplification may arise due to the existence of many parallel shortest paths (repeater chains) between two QPUs. However, a full network-aware quantum orchestrator in such network topologies is still fairly complex and out of scope of the current work.
We shall focus on switch-centric topologies when discussing the quantum orchestrator and presenting numerical simulations

We now briefly explain the modularity and scalability of server-centeric topologies using the BCube topology as an example.
This topology is deliberately designed for modularly scalable  data centers and can be expanded hierarchically to interconnect a large number of QPUs. For every expansion, the number of switches required at the added layer is jointly determined by the number of ports on switches and the number of layers in the network, e.g., there is one layer $k=1$ in Fig.~\ref{fig:arch}(b), and each switch has four ports $n=4$. BCube0 is obtained by connecting $n$ QPUs to an $n$-port switch. When building BCube1, an additional $n$ upper-layer switches are required. Each upper-layer switch is connected to all $n$ BCube0 containers, thereby constructing a larger BCube network recursively. The number of QPUs that a $k$-layer BCube can hold is $n^{k+1}$, at the cost of $k + 1$ ports at each QPU. There is a unique path between QPUs with the same index across the network, and there are at most $k + 1$ parallel paths (repeater chains) between any pair of QPUs in BCubek.

We present some additional server-centric topologies in Appendix~\ref{app:dcell} and summarize the number of network components in Table~\ref{tab:network-params}.

\section{Network-aware orchestrator for distributed quantum computing}
\label{sec:orchestrator}

\begin{figure}
    \centering
    \includegraphics[scale=0.6]{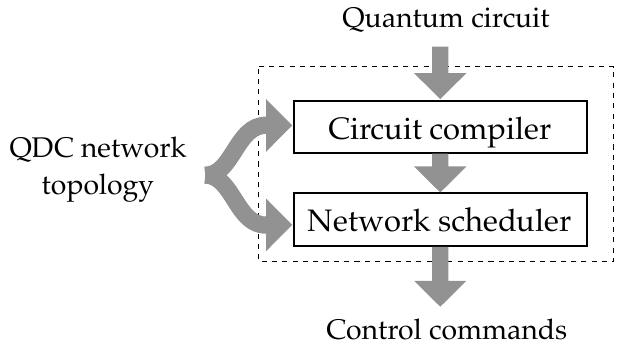}
    \caption{Pipeline for the proposed network-aware distributed quantum computing orchestrator.}
    \label{fig:qorchestra}
\end{figure}

Our focus in this paper is to lay foundations for the novel (physical layer) architectures for quantum data center networks. However, to have a working platform to perform distributed quantum computing jobs (in the form of quantum circuits of logical qubits), we see a need for several intermediate layers from ebit generation protocols on the physical layer all the way up to executing quantum applications. Along this line, the first step is to provide a framework for controlling the quantum hardware and switches to execute quantum jobs in a distributed manner.
In this section, we present the basic building blocks of such a framework which we call the quantum orchestrator. As shown in Fig.~\ref{fig:qorchestra}, this framework takes circuit-level description of quantum jobs as well as quantum network topology (including quantum hardware distribution across the network) as inputs and returns a set of instructions for optical switches (and other quantum hardware components). We note that these instructions are generated (offline) in advance, and are eventually executed by a classical central controller~\cite{dahlberg2019link,kozlowski2020designing,skrzypczyk2021architecture} to establish end-to-end ebits and consume them to execute remote gates.

We split the job of the quantum orchestrator into two steps: First, a circuit compilation where the circuit may be modified to be compatible with physical layer constraints, and ultimately logical qubits in the quantum circuit are mapped into physical qubits inside QPUs. Second, a network scheduling which goes through a given quantum circuit and finds a minimal sequence of commands for optical switching to perform remote gate execution based on the qubit mapping and available network resources. 

In what follows, we present a working version of the quantum orchestrator. The overall objective is to minimize the computation time and infidelity.
 It is worth noting that remote gates are generally slower and noisier compared to the local gates (see Sec.~\ref{sec:performance analysis} for some realistic numbers); hence, our objective is equivalent to reducing the number of remote gates, and among the remote gates selecting the less noisy ones (e.g., choosing the intra-rack communications over the inter-rack ones in switch-centric architectures). 
As we explain, each step of the quantum orchestrator may be broken down to several optimization sub-problems which require a more detailed analysis. We postpone the discussion on such details to future work. 

\subsection{Circuit compiler}
\label{sec:circuit_compiler}

The primary function of this module is to take an input quantum circuit of logical qubits and map these logical qubits to physical qubits within QPUs. In general, finding an optimal qubit mapping is an NP-hard problem~\cite{heunen2019automated}, and there is extensive literature on advanced algorithms for this purpose, including improvements over traditional graph partitioning methods like the KL and METIS algorithms~\cite{kernighan1970efficient,karypis1998fast}. 
In this paper, 
we consider a basic compiler with a static circuit partitioning algorithm; i.e., logical qubits are assigned at the beginning and inter-QPU gates are executed via gate teleportation. We
rely on heuristics to determine a qubit mapping that minimizes the number of remote gates, using standard graph partitioning techniques~\cite{kernighan1970efficient,karypis1998fast}. For simplicity, we assume that the input quantum circuit consists of only single-qubit and two-qubit gates; if not, it is transpiled into this form. While this assumption facilitates the use of graph partitioning algorithms, other circuit partitioning methods~\cite{heunen2019automated} exist that do not require it. We also assume that qubits inside the QPUs have all-to-all connectivity (as in trapped-ion or atomic based processors), otherwise, the compiler needs to modify the circuit to be compatible with the qubit connectivity patterns. Again, this is a standard step in circuit compilation and the existing tools can be adopted for this purpose. Besides that, quantum circuits can be designed based on the algorithm and the hardware constraints. This process is generally known as quantum architecture search (QAS)~\cite{martyniuk2024quantum}.
Our choice of compiler is simple but sub-optimal, and there are various ways for improvement such as allowing for qubit teleportation. This leads to adaptive circuit partitioning schemes~\cite{baker2020time,burt2024generalised} which will be detailed in a separate work~\cite{kaur2024optimized}.

For hybrid architectures that employ different protocols for generating ebits between QPUs located on the same rack and those on different racks, we address an additional optimization problem of assigning QPUs to racks. Specifically, we minimize an objective function, defined as a weighted sum of inter- and intra-rack ebits, with weights determined by the logarithms of their respective fidelities. As detailed in Appendix~\ref{app:rack assignment}, this optimization can be formulated as an integer linear program. Additionally, in Ref.~\cite{kaur2024optimized}, we extend the rack assignment problem to a more generalized setup involving an arbitrary number of link types.

\subsection{Network scheduler}

\begin{figure*}
    \centering
    \includegraphics[scale=.92]{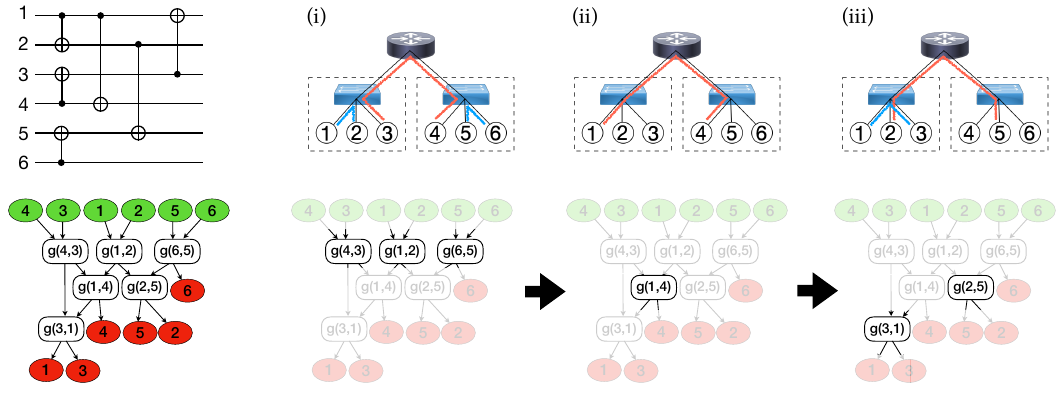}
    \caption{An instance of applying our scheduling algorithm to a quantum circuit of 6 qubits on a network of 6 QPUs. The circuit and its equivalent DAG are shown on the left. The other panels illustrate the step-by-step schedule of remote gates starting from the DAG frontier nodes. At each step, a set of remote gates to be executed are highlighted (while the rest are grayed out) and the corresponding switch configurations are shown as end-to-end paths. See the text for more details.}
    \label{fig:dag}
\end{figure*}

This module provides a set of commands, known as a network schedule, that sequentially controls quantum hardware and optical switches, managed by a central controller. Remote two-qubit gates are executed using the gate teleportation protocol~\cite{eisert2000optimal}, which consumes two ebits per remote gate. A key feature of this protocol is that the control qubit remains in its original QPU, unlike qubit teleportation, which modifies the qubit layout by moving the control qubit to the QPU containing the target qubit. Given that ebit generation is generally much slower than local gate execution, we ignore the local gate execution time, assuming it is effectively instantaneous. Here, we consider on-demand framework for switch-centric networks, where we generate ebits precisely when they are required for a specific remote gate. Further generalization to server-centric networks involve repeater network protocols since some QPUs cannot be directly connected. We imagine implementing repeater protocols within the data center based on asynchronous parallel protocols with entanglement swap as soon as possible~\cite{kamin2023exact,yang2024asynchronous,pouryousef2024minimal,goodenough2024noise}.

The scheduling commands encompass reconfiguring optical switches and allocating the necessary quantum network devices to establish an end-to-end optical path between the communication qubits. Multiple attempts are made to generate ebits until a heralding signal confirms success. Depending on the protocol, these commands require synchronization across devices, including the terminal QPUs, to ensure coordinated execution.

We first describe the scheduling algorithm for a single job scenario, i.e., scheduling for a quantum circuit. To create an effective schedule, we need to address two key issues: tracking gate dependencies (as gates acting on shared qubits may not commute) and optimizing for minimal switching events (to reduce latency) while maximizing network utilization by executing independent gates in parallel wherever possible. 

Gate dependencies are managed by applying a topological sorting algorithm to the quantum circuit computation graph, represented as a directed acyclic graph (DAG)~\cite{dagqiskit,hua2022exploiting}. We use a modified version of Kahn’s algorithm, designed to track dependencies while enhancing efficiency, as detailed below. The scheduling algorithm proceeds by iterating through three main steps until all gates in the DAG are scheduled. The set of commands are stored in the list $W_\text{total}$.

\textbf{Step 1:}  Identify the set of independent gates (also known as frontier nodes in the DAG) and add them to a set $I$. These are gates with no unexecuted dependencies, meaning they have no incoming edges in the current DAG state.

\textbf{Step 2:} Sort the independent gates in descending order based on the number of dependent gates (i.e., the number of successors in the DAG). Starting with the gate with the most dependencies, if the gate is remote, reconfigure the switches to establish an end-to-end path and allocate the required quantum devices along this path. For path selection, we always choose the shortest available path between two QPUs. Record the reconfiguration commands and allocated resources in $W$. Remove the corresponding node from the DAG. If a path or required resources are unavailable, proceed to the next gate in the list.

\textbf{Step 3:} Continue looping through Steps 1 and 2 as long as there are available optical paths and network resources. Once resources are exhausted, the list $W$ contains all commands for the current round of switching events. Add $W$ to $W_\text{total}$, clear $W$, and return to Step 1.

We note that in step 2 a heuristic approach is employed for resource management to mitigate routing and congestion. As detailed in Appendix \ref{app:dqc routing}, the complete resource management problem maps to a multi-commodity multi-flow problem, which can be formulated as an integer linear program. However, we opt for this heuristic method, which offers faster and more scalable performance for larger quantum jobs.

Figure~\ref{fig:dag} illustrates the above steps for a small quantum circuit on a network of two racks. Here, for clarity we show one qubit per QPU and assume there is one BSM per switch. The corresponding DAG is shown below the circuit. We use the same convention for DAG as Qiskit DAG~\cite{dagqiskit}, where qubits are shown as input and output vertices, gates are core vertices and edges connect input vertices to output vertices through the gates. 
Following the above steps, first it is evident from DAG that \textsc{cnot} gates acting on $(1,2)$, $(4,3)$, and $(6,5)$ are frontier nodes (here $(c,t)$ denote the control and target qubits of a \textsc{cnot} gate, respectively). We check that we have the resources to satisfy all these gates in parallel (step 2) and further see that we fully utilize the network resources (step 3). Hence, this forms our first switch configuration which is shown as (i) in the figure. Next, we go back to step 1 and look for a new set of independent gates and find that they are $(1,4)$ and $(2,5)$. However, we cannot execute both in parallel since there is only one BSM attached to the core (telecom) switch. Because gate $(1,4)$ has a successor in the DAG, we prioritize it and that becomes the next configuration (ii). Lastly, we are left with two independent gates $(2,5)$ and $(3,1)$ which can be executed in parallel. This forms the final switch configuration denoted as (iii).


Next, we discuss how to extend this algorithm to handle multi-job scheduling. Jobs may arrive at random times, including during the scheduling or execution of other jobs. In this approach, we do not assume prior knowledge of incoming jobs; instead, we learn about them as they are received. A different variant of multi-job scheduling arises in multi-tenancy scenarios, where we have a full list of jobs to be executed in the quantum data center and aim to find an optimal joint schedule that minimizes both latency and infidelity. We plan to address this latter problem in future work.

The first step in the multi-job scheduler is to maintain a job buffer, where we check if there are sufficient resources (in terms of available compute qubits) to accommodate an incoming job. If resources are available, we compile the circuit based on the unassigned QPUs and move the job to a scheduling list, where its DAG is added to the existing DAG being scheduled. If sufficient resources are not available, the job is placed in a waiting queue. This queue regularly checks for available QPUs and uses a first-in-first-out (FIFO) approach to assign pending jobs to free QPUs. If the waiting queue reaches capacity, the job is rejected.

In the multi-job setting, we extend Steps 1-3 from the above algorithm to operate across multiple DAGs, each associated with a different job. To ensure fairness, we introduce an additional condition in Step 2: we loop over all DAGs, scheduling one gate per DAG in each iteration. This approach balances the execution of tasks across multiple jobs.
This approach can further be improved to account for efficiency by prioritizing smaller jobs (with shallower circuit depth, fewer qubits, or both) when scheduled with large jobs so that they get executed faster instead of slowed down because of concurrency with those large jobs. 
Beyond that, there are numerous other opportunities for further optimization throughout the compilation and scheduling stages, which we plan to explore in future work.

\section{Performance analysis}
\label{sec:performance analysis}

In this section, 
 we present some simulation results where we combine physical layer modeling, network protocols, and network-aware orchestrator. For the QDC architecture, we focus on the Clos topology with hybrid protocols for intra- and inter-rack quantum communications.
Because of the probabilistic nature of entanglement generation protocols, we use average quantities to estimate 
average network latency, i.e., how long the circuit execution takes on average, and a proxy for circuit fidelity, i.e., how noisy the platform is. Although these two quantities, namely, rate and fidelity (or their variants), can be combined to define a quantum network utility function~\cite{vardoyan2022quantum,lee2022quantum}; we focus on addressing them separately in this paper.

As mentioned in the previous section, a network schedule is a list of switching events. Each switching event consists of a set of ebits to be generated in parallel. Let $n_r$ be the  number of pairs of QPUs which need to generate ebits at $r$-th round of switching and $t_i$ denote the time it takes to generate an ebit for $i$-th QPU pair with $i=1,2,\cdots, n_r$. Hence, the duration of this switching event is given by
\begin{align}
    \label{eq:t-round}
    T_r = \max(t_1, t_2, \cdots, t_{n_r}).
\end{align}
where $t_i$ is a random number which is simply related to the number of attempts until success and whose probability distribution depends on the ebit generation protocol (c.f.~Sec.~\ref{sec:Entanglement generation process}). The overall execution time is then found by aggregating the duration of switching rounds
\begin{align}
    T_\text{tot} = \sum_r (\tau_\text{sw} + T_r),
\end{align}
with an additional contribution due to the reconfiguration time $\tau_\text{sw}$ of optical switches in each round. In our numerical evaluations, we compute an estimate for the expectation value of each round duration~Eq.~(\ref{eq:t-round}) using Monte-Carlo, i.e., by sampling from the distribution of each $t_i$ and aggregate the result over many iterations.







We also calculate a weighted sum of number of gates as a proxy for the quality of distributed quantum computation. Concretely, for $M$ types of gates (including local and various non-local types) we define a cost function as follows
\begin{align}
    \label{eq:c-fid}
    C_\text{Fid} = \sum_{i=1}^{M} n_i \left(\frac{\log F_i}{\log F_1}\right),
\end{align}
where $n_i$ is the number of $i$-th type gates and $F_i$ denotes the respective average fidelity of these gates. We choose one of the gate types as a baseline and work with the ratios of log fidelities which do not depend on log basis anymore. It is important to note that the quantity (\ref{eq:c-fid}) is a measure of infidelity, i.e., the smaller the better (see Appendix~\ref{app:rack assignment} for derivation details). In our analysis we only consider two gate types intra- and inter-rack gates and this expression is simplified into
\begin{align}
    \label{eq:logfid-hybrid}
    C_\text{Fid}^{\text{(hyb)}} = n_\text{loc} + n_\text{intra} \left(\frac{\log F_\text{intra}}{\log F_\text{loc}}\right) + n_\text{inter} \left(\frac{\log F_\text{inter}}{\log F_\text{loc}}\right).
\end{align}

\begin{figure}
    \centering
    \includegraphics[scale=0.9]{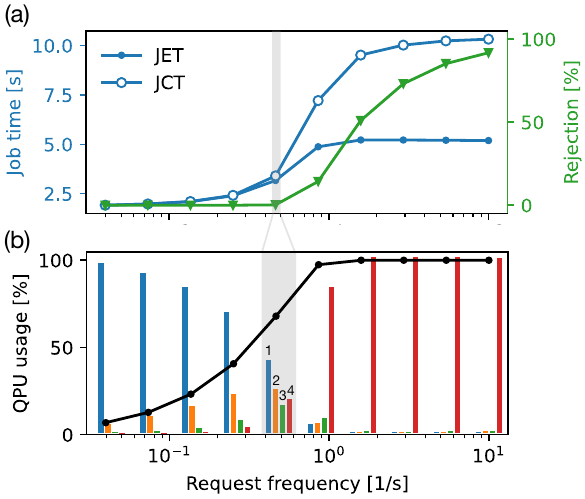}
    \caption{Multi-job scheduling for jobs of identical size. (a) Job times (blue open and close circles) and rejection rates (green triangles), and (b) networked QPU usage (black dots) for a series of random jobs of equal size which arrive at random times according to a Poisson process (average rate is the request frequency). For each value of request frequency, a bar plot in (b)  demonstrates the percentage of $n$-QPU usage per rack on average. This process is illustrated for one of the middle data points where the color coding for $n$ is also indicated. See Table~\ref{tab:random-bench} for network parameters. }
    \label{fig:identical-jobs}
\end{figure}


We now present the numerical values we plug in to our numerical analysis. We consider emitter-emitter protocol with Fock space encoding for intra-rack communications as explained in Sec.~\ref{sec:emitter-emitter}.
Considering hardware parameters from Refs.~\cite{monroe2014large,saha2023low,zhou2024long,stephenson2020high} $\alpha=0.05$, $\eta=0.1$ (i.e., 10 dB loss), and $\tau_0^{-1}\sim 1$ MHz, we obtain $\tau_\text{ee}= 0.1$ ms and $F_\text{ee} = 0.95$ for the same rack EPR pair generation. 
For the inter-rack communication, we use scatterer-scatterer protocol where the end-to-end ebit generation rate depends on various hardware parameters as discussed in Sec.~\ref{sec:scatterer-ccatterer}. With some reasonable parameters such as $10^6$ end-to-end photon pair generation of entanglement source with $1$GHz photon linewidth, and $1\mu$sec qubit reset time, the exponential distribution parameter becomes $\lambda_{ss}^{-1} = 10$msec. We also consider the average optical switch reconfiguration time to be $\tau_\text{sw}=1$msec. These numbers are typical values for off-the-shelf devices and do not necessarily represent the state-of-the-art devices.

In what follows, we consider two representative examples: 
\begin{itemize}
    \item 
    To demonstrate how network scheduler handles multi-job scenarios, we consider a stochastic sequence of random quantum circuits of varying size and depth. We construct random quantum circuits  of $n$ qubits with a square form factor (equal width and depth).
    
    \item 
    As a second example, we evaluate the network latency and compute infidelity for some common quantum algorithms which are often used as subroutines for more complex algorithms. 
\end{itemize}

\subsection{Random benchmarks}
\label{sec:random-bench}

In this experiment, we consider quantum jobs arriving at random times according to a Poisson distribution which is characterized by the number of jobs per unit time, $\gamma$, called request frequency. In other words, the probability of having $k$ jobs over a period $t$ is given by $\frac{(\gamma t)^k}{k!} e^{-\gamma t}$. Our goal here is to generate network traffic (synthetically) and study how our design performs in terms of network latency in various scenarios. To this end, we consider random jobs in the form of random quantum circuits.
We skip the compiler step for these jobs since such quantum circuits are featureless, i.e., all two-qubit gates are equally likely, and our simple qubit mapping algorithm will not have a a significant impact on average.

We note that since we have a job queuing pipeline before sending jobs for the execution. Hence, we define two relevant quantities (which we collectively call job times) to capture the impact of network latency and multi-job execution:
\\
\textbf{Job completion time (JCT):} Time difference between the moment the job is finished and the moment the job is submitted.\\
\textbf{Job execution time (JET):} Time takes to execute the quantum circuit from the moment the computation starts.

Clearly, JCT for an accepted job is equal to the JET plus the buffer time.

Table~\ref{tab:random-bench} summarizes the size of the network and number of various components used for simulations in this part.

\begin{table}[]
    \centering
    \begin{tabular}{l|c|c|c}
    \Xhline{2\arrayrulewidth}
   Component &  Figs.\ref{fig:identical-jobs}, \ref{fig:different-jobs}(a) & Figs.\ref{fig:different-jobs}(b),(c) & Table~\ref{tab:algo-bench} \\
    \Xhline{2\arrayrulewidth}
    Switches  & 18 & 28 & 10 \\
    BSMs per switch  & 4 & 4 & 2\\
    Racks &  9 & 16 & 4 \\
    QPUs per rack &  4 & 8 & 4 \\
    QPUs  &  36 & 128 & 16 \\
    Data qubits per QPU  &  10 & 10 & 20 \\
    Comm. qubits per QPU & 4 & 4 & 4 \\
    \Xhline{2\arrayrulewidth}
    \end{tabular}
    \caption{Number of various components in the quantum data center network with Clos topology $n=6$  studied in Secs.~\ref{sec:random-bench} (Figs.~\ref{fig:identical-jobs} and \ref{fig:different-jobs}) and \ref{sec:algo-bench} (Table~\ref{tab:algo-bench}).}
    \label{tab:random-bench}
\end{table}



\subsubsection{Equal job sizes}

In this experiment, we consider a sequence of random quantum circuits each with the same number of qubits such that each job is distributed across all racks utilizing one QPU per rack. Our motivation for coming up with such a choice is to create a synthetic traffic and congestion in the telecom layer of the Clos network in a controlled way so that we can interpret/understand the results easily.
 The number of various components are summarized in Table~\ref{tab:random-bench}. In particular, each job takes $90$ qubits, and the circuit depth is equal to the circuit width. We run these jobs on a Clos network with $9$ racks, and each rack has $4$ QPUs. There are $10$ data qubits per QPU. Hence, by design each job requires $9$ QPUs which are distributed over all racks, and maximum number of parallel jobs is $4$. We assume the job queue buffer size is $4$, i.e., we only keep $4$ jobs in the queue before sending them to the scheduling module; more jobs arriving will be rejected.

Figure~\ref{fig:identical-jobs} shows how the performance metrics evolve as we increase the job arrival rate (or request frequency) from fewer than 1 job every 10 seconds to 10 jobs per second. For every request frequency, we compute JET and JCT, rejection rate, and the QPU usage as defined by
\begin{align}
    \label{eq:qpu-formula}
    \text{QPU usage} = \frac{\text{Total QPU time usage}}{\text{Total No. of QPUs}\times \text{JET}},
\end{align}
which is reported in percentage and used to capture how much of the QDC compute power is utilized for a given request frequency. We further show a bar plot for each data point which illustrates QPU-resolved per rack usage i.e., out of the QPU percentage usage how much of it (in percentage) is $1$ QPU per rack, $2$ QPU per rack, etc. 

To obtain each data point in Fig.~\ref{fig:identical-jobs}, we averaged these quantities over $10^3$ iterations where each iteration is a time interval long enough that it contains $10^4$ requests on average. 
As we move from left to right along the horizontal axis, when the request frequency is too small, we are at the single job scheduling regime where a given job finishes execution before the next job arrives. This results in identical values for the JET and JCT since there is no queuing time, which is about $2.3$sec as we see in Fig.~\ref{fig:identical-jobs}(a). This is also evident in Fig.~\ref{fig:identical-jobs}(b), since QPU usage is very low (network is idle most of the time) and when in use it is nearly $100$\% of time operating at $1$ QPU per rack (c.f. blue bars (b)). As we move past $\gamma = 4\text{sec}^{-1}$ (corresponding to $4$ jobs arriving every second), we start to have more overlapping jobs (i.e., we use more QPUs per rack as seen in bars other than blue), rising job times, and JCT and JET bifurcating (as queue is being formed). For higher request frequencies, then we utilize the entire resources (also $4$ QPU per rack (red bar) is nearly $100$\%) and JET asymptotically reaches the four parallel job regime $5.1$sec which is nearly twice as long compared to the single-job regime. Another way to see the job times plateauing is that QDC can only handle $4$ jobs and the rest are sent to the queue. At the same time, because queue buffer has the capacity for 4 jobs then most requests are rejected.

\subsubsection{Variable job sizes}

Here, we present the results of two numerical experiments where random quantum circuits with different number of qubits arrive at various rates. In either experiments, our policy for QPU assignment in the waiting queue is FIFO provided that we meet the required number of QPUs, otherwise we check the subsequent jobs in the queue. Figure~\ref{fig:identical-jobs} summarizes the simulation results, where each data point is obtained by averaging over $10^3$ iterations where each iteration is a time interval long enough that it contains $10^4$ requests on average.

\begin{figure}
    \centering
    \includegraphics[scale=0.72]{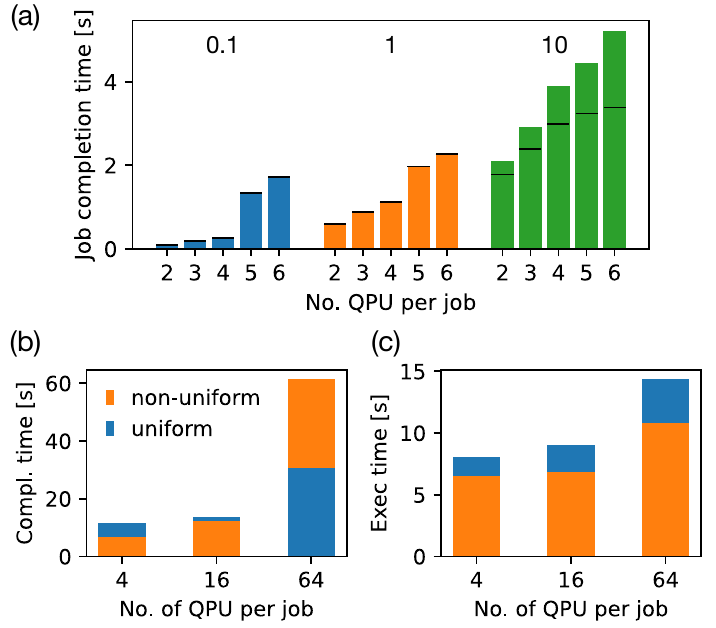}
    \caption{Job completion times for multi-job scheduling of variable job sizes. In (a), the job sizes are incremental, and 
    in (b) and (c) there are three job sizes of 4, 16, and 64 QPUs, which represent small, intermediate, and large job sizes, respectively. Horizontal black lines in (a) indicate the corresponding job execution times.}
    \label{fig:different-jobs}
\end{figure}

As a first experiment, we consider jobs of different sizes in an incremental order each requiring $n = 2, 3, \cdots, 6$ QPUs, respectively, arriving at the same rate. We show the job times for three request frequencies $0.1$sec$^{-1}$ (i.e., one job every $10$sec on average), $1$sec$^{-1}$ (i.e., one job every second on average), and $10$sec$^{-1}$ (i.e., 10 jobs every second on average) in Fig.~\ref{fig:different-jobs}(a). 
For slow request frequencies the average QPU occupancy is $\sum_n=20$ which is well below the total number of QPUs (c.f.~Table~\ref{tab:random-bench}) and the job arrival times are well separated. Hence, the waiting queue almost always remains empty, and there is no difference between JET and JCT. 
We note that there is a clear jump in the JET as we go from $4$(and below)-QPU jobs to $5$(and above)-QPU jobs. The reason is jobs with $4$ or fewer QPUs are placed on the same rack and ebit generation is exclusively intra-rack at NIR frequencies, which are much faster than the inter-rack (telecom) ebit generation processes.

The small difference between JCT and 
JET also holds for $\gamma = 1$sec$^{-1}$, since the slowest job is the one requiring $6$ QPUs (which takes around $2$sec to finish) and the network can accommodate two $6$-QPU jobs (since during this $2$sec on average two jobs of this type arrive). That said, we still observe an increase in job completion time due to more jobs running in parallel and sharing the network resources. 

When $\gamma = 10$sec$^{-1}$, we see that a large gap between JCT and JET develops as the queue is filling up. Aside from that, after a few rounds of QPU assignments, there will be random QPU availability across the racks. For example, there will be two or three empty QPUs which may not belong to the same rack and assigned to $2$- or $3$-QPU jobs. This leads to significantly longer execution times ($T_\text{tot}\sim 2$sec) for such jobs compared to the the case with slower request frequencies ($T_\text{tot} \sim 100$msec), where small jobs are mostly placed on the same rack. The reason is that quantum communications now run at telecom (inter-rack) which are significantly slower than the intra-rack communications.

One may ask what would happen if there is a huge disparity between the job sizes. A typical scenario would be when large jobs are requested at slower rates and small jobs are requested at higher rates. This is studied in  Fig.~\ref{fig:different-jobs}(b) and (c). Here, we consider three job sizes in the form of random quantum circuits requesting $4$, $16$, and $64$ QPUs representing small, intermediate, and large jobs which are requested at different rates $10$, $1$, and $0.1$ per sec, respectively. The data center is large enough to be able to accommodate two large jobs at the same time (c.f.~Table~\ref{tab:random-bench}).
For reference, we also show the corresponding values of JET and JCT when all job types arrive at the same rate $\gamma = 1$sec$^{-1}$. A (rather surprising) overall observation in Figs.~\ref{fig:different-jobs}(b) and (c) is that for small and intermediate jobs both JCT and JET are larger for uniform rates, while the order changes for the large jobs. We give an explanation below.

The hierarchy in JCT is evident for non-uniform rates (orange bars) in Fig.~\ref{fig:different-jobs}(b) as there are more small or intermediate jobs in the queue (due to larger request frequencies) which will more likely be assigned faster (since they need fewer QPUs); however, we need to wait long enough until half of total QPUs become available for a large job. In contrast, when all job types have the same request frequency, we have similar number of different job sizes in the queue and the wait time for $64$-QPU job drops significantly because they can replace an old $64$-job which just finish executing. At the same time, there is a rise in the JCT of small and intermediate job sizes, i.e., they are kept longer in the queue, since they may not bypass a large job in the queue anymore. 

In contrast, the variation in JET for different job sizes (as shown in Fig.~\ref{fig:different-jobs}(c)) is not as much as that of JCT. An important observation is that although $4$-QPU jobs can fit in a rack (which hosts $8$ QPUs) JET of $4$- and $16$-QPU jobs are fairly close. This is because $4$-QPU jobs often share a rack with $16$-QPU jobs and our scheduler distribute resources evenly between different jobs. As a result, $4$-QPU job execution is slowed down by the neighboring $16$-QPU job (which require intra-rack communications) with which it is sharing a rack. Compared to the uniform request rate, JETs are generally smaller when the rates are non-uniform because in the former case the odds of sharing resources with similar or greater jobs are higher which leads to an increased JET on average.




\subsection{Algorithmic benchmarks}
\label{sec:algo-bench}

\begin{table}
    \centering
    {\renewcommand{\arraystretch}{1.3}
    \begin{tabular}{@{\extracolsep{9pt}}p{0.12\textwidth}>
    {}p{0.045\textwidth}>
    {}p{0.045\textwidth}>{}p{0.05\textwidth}>{}p{0.045\textwidth}>{}p{0.05\textwidth}}
    \Xhline{2\arrayrulewidth}
&{}{}   
       &  \multicolumn{2}{l}{Our compiler}  &  \multicolumn{2}{l}{Baseline}
    \\
   \cline{3-4}  \cline{5-6} 
   Circuit & $\cnot{}{}$ gates &Infid  & Time[s] & Infid & Time[s] \\
   \Xhline{2\arrayrulewidth}
    BV (280)  &152& 1505 & 0.47 & 2130 & 0.71  \\
    QFT (100)  &6510& 39470 & 20.31 & 68191 & 27.65  \\
    QV (100)  &15000& 107155 & 35.89 & 158242 & 40.10  \\
    ISING (98)  &194& 248 & 0.029 & 2014 & .42  \\
    ADDER (118)  &845& 1236 & 0.54 & 9624 & 3.09  \\
    CAT (260)  &259& 345 & 0.033 & 3164 & .74  \\
    Swap (115)  &456& 2117 & 0.86 & 5716 & 1.97  \\
    \Xhline{2\arrayrulewidth}
    \end{tabular}
    }
    \caption{Performance metrics for algorithmic quantum circuits from \cite{li2023qasmbench}. Baseline is averaged quantities for random qubit/QPU assignments (see main text for details). The numbers in parenthesis denote the number of logical qubits in the circuit. Network parameters are provided in Table~\ref{tab:random-bench}. 
    \label{tab:algo-bench}}
\end{table}

We evaluate the proposed circuit compiler in Section~\ref{sec:circuit_compiler}, which incorporates qubit assignment via graph partitioning and QPU assignment to racks using Integer Linear Programming (ILP), against a baseline of random qubit and QPU assignments. The comparison is conducted on several well-known benchmark circuits (studied in Ref.~\cite{li2023qasmbench}), a Quantum Fourier Transform (QFT) circuit, and a Quantum Volume circuit for 100 qubits.

Performance is assessed using two key metrics: Eq.~(\ref{eq:logfid-hybrid}) and job execution time, with results summarized in Table~\ref{tab:algo-bench}. For random assignment, the results are averaged over 100 independent runs of the compiler. Overall, our analysis demonstrates that the proposed circuit-compiler consistently outperforms random assignment across all benchmarks and metrics, highlighting its effectiveness in optimizing quantum circuit execution.

In Table~\ref{tab:algo-bench}, we observe the following trend: in general, the infidelity increases as the number of $\cnot{}{}$ gates increases. This can be explained as follows: even in the absence of non-local gates, the infidelity would still scale with the number of $\cnot{}{}$ gates, as indicated in Eq.~(\ref{eq:logfid-hybrid}).

Examining the trends further, consider the BV(280) and ISING(98) circuits. Although the number of $\cnot{}{}$ gates in these circuits is approximately the same, the observed infidelity differs significantly. This discrepancy can be attributed to the number of non-local $\cnot{}{}$ gates, which is higher in the BV(280) circuit. In summary, the observed infidelity is influenced by two key factors: the total number of $\cnot{}{}$ gates and the fraction of those gates that are non-local.


\section{Discussion}
\label{sec:discussion}

In conclusion, this work introduces scalable architectures for quantum data center networks based on dynamic circuit-switched quantum networks that distribute entanglement between QPUs.
 By utilizing shared quantum resources and adopting modular topologies, such as switch-centric and server-centric designs, we achieve on-demand, all-to-all connectivity while minimizing reliance on costly quantum hardware.
 We developed a network-aware quantum orchestrator and entanglement generation protocols to manage distributed quantum computing jobs, connecting physical-layer architectures with quantum applications. 
 Through simulations and benchmarking, we evaluate the circuit execution capabilities of our architectures, demonstrating the opportunities and challenges in scalability, efficiency, and fidelity.

Our research establishes a foundation for the development of large-scale quantum computing infrastructure, bringing us closer to achieving practical quantum advantage. We have introduced novel quantum data center network architectures that go beyond traditional peer-to-peer designs, opening up several avenues for future research.
These architectures are not mutually exclusive but rather complementary, allowing for the combination of their strengths. For example, the inter-rack topology can be server-centric (as opposed to a star topology with a top-of-rack switch), while different racks can be connected using a switch-centric network.
We have not yet explored the integration of quantum memories with optical switches in the network, which could offer more flexibility in terms of time synchronization and improved rates for seamless entanglement generation. Some initial efforts in this direction have been made by Choi~\emph{et al.}~\cite{choi2023scalable}, who incorporated quantum memories into a fat-tree network topology.


Throughout our paper, we adopt model abstractions and intentionally avoid delving deeply into the specifics of physical qubits, though our models are primarily inspired by cold atoms and trapped ions. While quantum hardware remains in its early stages and requires significant advancements to enable true large-scale quantum data centers, significant progress is being made. For instance, recent developments in photonic quantum technologies, such as nanofiber-based optical cavities serving as spin-photon interfaces, are laying the groundwork for scalable quantum data centers~\cite{sunami2024scalable}. 

Moreover, while superconducting qubits have traditionally been the cornerstone for monolithic systems, notable progress has been achieved in distributed quantum computing between superconducting processors~\cite{ang2024arquin}. For example, our hybrid architecture could incorporate microwave-to-telecom transducers, such as those developed in Ref.~\cite{mayor2024two}. Although current ebit generation rates with these transducers are quite low (around a few hundredths of a hertz), their integration demonstrates the feasibility of bridging various quantum platforms, highlighting the potential for future advancements.

In our simulations, we accounted only for gate infidelity, neglecting the impact of qubit coherence time. While this approach is reasonable for highlighting that remote operations are significantly more error-prone than local quantum gates by assigning them lower fidelities, it overlooks a critical factor: data qubits used in quantum computation decohere over time. The extended duration of quantum communication between QPUs can significantly degrade computational quality due to qubit decoherence. This effect has not been included in our simulations. As shown in Sec.~\ref{sec:performance analysis}, the computation time can span several seconds, which is comparable to the coherence time of various quantum technologies, particularly cold atoms and trapped ions. Consequently, it is essential to incorporate coherence time considerations into future analyses.

In this paper, we focused on a near-term application where computations are performed on individual qubits. However, an additional intermediate layer could be introduced to the QDC network stack to enable entanglement distillation or, more generally, quantum error correction. In our study, we used `bare' ebits as they were generated, without applying purification or error correction. While techniques like distillation or error correction can enhance fidelity, they incur additional costs in terms of time—leading to increased decoherence—and require more hardware resources. With quantum error correction, the network architecture may need to be adapted to include QPUs that utilize logical qubits (e.g., surface code). In such a setup, generating a logical ebit could involve producing \(d\) physical ebits for a logical qubit encoded using a surface code of distance \(d\)~\cite{ramette2024fault,sinclair2024fault}. In this context, exploring the trade-offs between simplicity and time efficiency in various ebit generation protocols, such as sequential versus parallel ebit generation, would be an interesting avenue for future research.

To enable distributed quantum computing, we introduced the basic components of a quantum orchestrator designed to control quantum hardware and switches for executing jobs across a network. Our approach included a basic compiler with static circuit partitioning, where qubits are assigned at the outset, and inter-QPU gates are executed via gate teleportation. Future enhancements could involve advanced compilers with adaptive circuit partitioning~\cite{baker2020time,burt2024generalised,kaur2024optimized}.
The current circuit compiler necessitates an ebit count proportional to the number of remote gates, but number of necessary ebits can be reduced through compiler-based communication fusion~\cite{wu2023qucomm,wu2022autocomm}. 
While our orchestrator employs sequential optimization steps, combining these into a holistic end-to-end optimization framework with feedback loops could offer greater efficiency. Such a framework would simultaneously consider qubit mapping, classical control signals, and other factors, likely resulting in complex nonlinear optimization problems. These challenges could be addressed with intelligent methods like reinforcement learning or genetic algorithms~\cite{crampton2024genetic,promponas2024compiler}, though these approaches are computationally intensive. Thus, developing effective heuristics remains a critical area for future work.


In our numerical simulations, we primarily used random quantum circuits to remain agnostic to specific algorithm details. Specifically, we employed a Poisson process to model multi-job scheduling in our network orchestrator, simulating generic network traffic composed of random circuits. In the future, it would be valuable to define and study benchmarks or network traffic by categorizing quantum circuits based on their structural characteristics and utilizing representative benchmark sets derived from clustering similarly structured circuits~\cite{bandic2024profiling}.


Finally, the idea of integrating QPUs as accelerators within HPC infrastructures has gained significant attention recently~\cite{mohseni2024build}. Rather than replacing classical computers as general-purpose systems, quantum computers can excel at specialized tasks. Exploring the impact of our work on designing modular hybrid architectures—comprising multiple QPUs interconnected with classical HPC nodes—presents an exciting avenue for future research. On the orchestrator side, further advancements are needed to support seamless quantum-classical integration, potentially leading to the development of future middleware systems~\cite{saurabh2023conceptual}.

\acknowledgements

The authors acknowledge insightful discussions with Stephen DiAdamo, Don Towsley, Yufei Ding, Luca Della Chiesa, Galan Moody, and Raj Jain. 

\appendix



\section{Simulations of scatterer-scatterer protocol}
\label{app:scatter-scatter}

In this appendix, we numerically simulate the scatterer-scatterer protocol with probabilistic sources and present the statistics of success times.

Our starting point is that an incoming photon wavepacket with mean characteristic time $t_i$ is described by a Gaussian envelope function with the central frequency $\omega_0$ and  linewidth of $\Delta \omega$
\begin{align}
    \label{eq:photon-packet}
    \ket{t_i}_{a_i} = a_{i,t_i}^\dag \vac \equiv \int d\omega \frac{e^{-\frac{(\omega-\omega_0)^2}{2\Delta\omega^2}}}{\pi^{1/4} \Delta\omega^{1/2}}  e^{i\omega t} a_\omega^\dag \vac
\end{align}
where $i$ is to denote the $i$-th photons (could be presense/absence or time-bin).

We now explain how we carry out the simulations step by step. A pseudo-code of our algorithm is given in Algorithm~\ref{alg:ebit-gen-sim}.
We generate two random sequence of photon pair generation events associated with the two sources according to the Poisson distribution parameter $\lambda$. Next, we find the pairs which are one after another and check if communication qubits are available (i.e., they are not undergoing reinitialization). If yes, we accept this event with probability
\begin{align}
    \label{eq:overlap}
    p_\text{ss} = \frac{1}{2} |\!\braket{t_2|t_1}\!|^2 = \frac{1}{2}  e^{-\frac{1}{2}\Delta\omega^2(t_1-t_2)^2},
\end{align}
or reject with probability $1-p_\text{ss}$ in which case the communication qubits go through a reinitialization (and will not be available for a duration of $\tau_0$). Here, the factor of $1/2$ is to account for the BSM success probability.
Therefore, we observe that there is some finite probability for the coincidence as long as $|t_1-t_2| \Delta \omega \lesssim 1$.

\begin{figure}
    \centering
    \includegraphics[width=.8\linewidth]{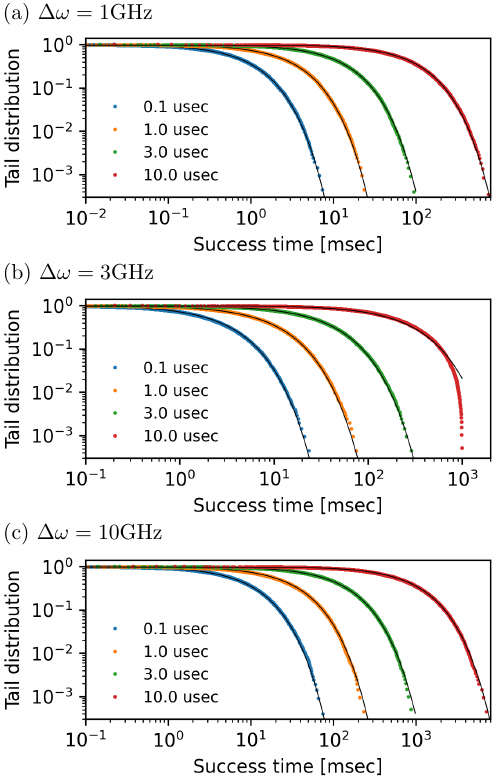}
    \caption{Tail distribution function of success time for scatterer-scatterer protocol for various qubit reinitialization times shown in the legend. Black solid lines are the fit according to the exponential distribution $\text{Pr}(T>t) =e^{-\lambda_\text{ss}t}$, (c.f.~Eq.~(\ref{eq:exp-dist})).}
    \label{fig:exp-dist}
\end{figure}

The above algorithm constitutes one iteration, and we must repeat this many times to accumulate some statistics. In Fig.~\ref{fig:exp-dist}, we show the tail distribution function (or complementary cumulative distribution function $\text{Pr}(X> x) = 1-\text{Pr}(X\leq x)$) of successful events after running $10^5$ iterations where we only include iterations which end up with a successful events (since we are interested in the statistics of success time).

\begin{figure}
    \centering
    \includegraphics[width=1.01\linewidth]{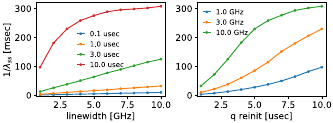}
    \caption{The fitted exponential distribution parameter as a function of photon linewidth and qubit reinitialization time. Legend in left (right) panel denotes the values of qubit reinitialization time (photon linewidth). }
    \label{fig:exp-dist-param}
\end{figure}

We observe that the tail distribution aligns well with the exponential distribution described in Eq.~(\ref{eq:exp-dist}), where the parameter $\lambda_\text{ss}$ depends on system characteristics such as photon linewidth and qubit reinitialization time, as illustrated in Fig.~\ref{fig:exp-dist-param}. Generally, longer qubit reinitialization times or larger photon linewidths result in smaller rates $\lambda_\text{ss}$, leading to longer entanglement generation times. This is because entangled photons cannot be stored in communication qubits during reinitialization, and larger photon linewidths produce shorter photon wave packets with a reduced probability of overlap, as implied by Eq.~(\ref{eq:overlap}).


\begin{algo}{Simulating Entanglement generation protocol} 
\Input{Two lists of emission events $P_1$ and $P_2$.} 
\Output{Determine whether end-to-end entanglement is realized or not, $\text{ebit}={\bf True}/{\bf False}$.}
\label{alg:ebit-gen-sim}
    Combine $P_1$ and $P_2$ and sort them, $P={\bf sorted}(P_1+P_2)$.\;
    Define control variables ${\bf bool}\ R_1, R_2 := {\bf False}$ whether comm. qubits are undergoing resetting or not.\;
    Define time when comm. qubits are available after resetting ${\bf float}\ T_1, T_2 := 0$.\;
    Initialize $\text{ebit}={\bf False}$.\;
    \For{$p_i, s_i \in P$}{
        {\it ($p_i$, photon emission time, $s_i$, photon source)}\;
        {\bf If} {$p_i \geq T_{s_i}$}
        {\bf then} $Q_{s_i} = {\bf False}$, $T_{s_i}=0$.\;
        \uIf{$s_i = \overline{s_{i+1}}$ (two consecutive photons belong to two sources)}{
            {\bf If} {$p_{i+1} \geq T_{s_{i+1}}$}
            {\bf then} $Q_{s_{i+1}} = {\bf False}$, $T_{s_{i+1}}=0$.\;
            \If{$Q_{s_i} = Q_{s_{i+1}} = {\bf False}$ }{
                $\Delta t = p_{i+1}-p_i$.\;
                Accept ebit with probability ${\bf overlap}(\Delta t)$, $\text{ebit}={\bf True}$, and {\bf exit}.\;
                }            
            }
        {\bf If} {$p_i \geq T_{s_i}$}
        {\bf then} $Q_{s_i} = {\bf True}$, $T_{s_i}=p_i + T_\text{reset}$.\;
    }
    {\bf return} $\text{ebit}$.
\end{algo}

\section{Additional network topologies}
\label{app:additional topologies}

In this section, we present a few more network topologies for each category as explained below. The number of network switches, QPUs, and the network diameter are shown in Table~\ref{tab:network-params} for reference~\cite{liu2013data}. 

\begin{table*}[]
    \centering
    {\renewcommand{\arraystretch}{1.3}
    \begin{tabular}{@{\extracolsep{20pt}}p{0.12\textwidth}>{}p{0.1\textwidth}>{}p{0.1\textwidth}>{}p{0.12\textwidth}>{}p{0.12\textwidth}>{\arraybackslash}p{0.1\textwidth}}
    \Xhline{2\arrayrulewidth}
       &  \multicolumn{2}{l}{Switch-centric}  &  & \multicolumn{2}{l}{Server-centric} \\
       \cline{2-4}  \cline{5-6} 
       & Basic tree & Fat-tree & Clos network & DCell & BCube \\
       \Xhline{2\arrayrulewidth}
      Diameter &  $2\log_{n-1}N$ & 6 & 6 & $2^{k+1}-1$ & $\log_n N$ \\
      No. of switches &  $\frac{n^2+n+1}{n^3} N$ & $\frac{5N}{n}$ & $\frac{3}{2}n+\frac{n^2}{4}$ & $\frac{N}{n}$ & $\frac{N}{n}\log_n N$ \\
      No. of QPUs &  $(n-1)^3$ & $\frac{n^3}{4}$ & $\frac{n^2}{4}\times n_\text{ToR}$ & \makecell[l]{$\geq (n+\frac{1}{2})^{2^k} -\frac{1}{2}$ \\  $\leq (n+1)^{2^k} -\frac{1}{2}$ }
      & $n^{k+1}$ \\
    \Xhline{2\arrayrulewidth}
    \end{tabular}
    }
    \caption{Summary of network parameters. $n$ is the number of outgoing ports of each switch  to other switches or QPUs. For $n_\text{ToR}$ specifically for Clos network denotes the number of QPUs per rack. $N$ is the total number of QPUs. Diameter is defined by the longest of shortest paths between  QPUs. $k$ is the number of layers (or levels) in a server-centric network.}
    \label{tab:network-params}
\end{table*}

\subsection{Switch-centric topology}
\label{app:hyperx}


A natural extension of the star topology (e.g., a top-of-rack switch) is a simple tree structure, where switches are connected hierarchically, as illustrated in Fig.~\ref{fig:network-app}(a). However, this connectivity creates a performance bottleneck and introduces a single point of failure at the core level. To address these issues, the Fat-Tree topology~\cite{al2008scalable} was proposed to enable non-blocking transmission. 

In a Fat-Tree topology, the links connecting nodes in adjacent tiers become progressively wider as they ascend the tree toward the root. This is accomplished by designing the topology such that, for any switch, the number of links connecting to its children equals the number of links connecting to its parent, assuming all links have the same capacity. In this configuration, each \(n\)-port switch in the edge tier is connected to $n/2$ servers, while the remaining $n/2$ ports link to \(n/2\) switches in the aggregation tier. Together, the $n/2$ aggregation switches, $n/2$ edge switches, and the servers form a basic unit of the Fat-Tree, referred to as a pod. At the core level, there are $(n/2)^2$ $n$-port switches, each connecting to all \(n\) pods. 

Figure~\ref{fig:network-app}(b) depicts a Fat-Tree topology with \(n=4\). Unlike the simple tree topology, the same type of switches is used across all three levels of the Fat-Tree. Moreover, high-performance switches are not required in the aggregation and core levels, making the Fat-Tree an efficient and scalable solution.


 A HyperX network~\cite{6375529} is a direct network of switches, where each switch connects to a fixed number \(T\) of terminals. In general, a terminal can represent a compute node, a cluster of compute nodes, or any other interconnected device. In our case, we use terminals as top-of-rack (ToR) NIR switches. The switches are represented as points in an \(L\)-dimensional integer lattice, with each switch uniquely identified by a coordinate vector \(I = (I_1, \dots, I_L)\), where \(0 \leq I_k < S_k\) for each \(k = 1, \dots, L\). 

Within each dimension, the switches are fully interconnected. Consequently, each switch has bidirectional links to exactly \(\sum_{k=1}^L (S_k - 1)\) other switches, connecting to all other switches that differ in only one coordinate. The total number of switches \(P\) in the HyperX network satisfies \(P = \prod_{k=1}^L S_k\). A regular HyperX network is defined by the condition \(S_k = S\) for all \(k\), and is characterized by the tuple \((L, S, K, T)\). 

Figure~\ref{fig:network-app}(c) illustrates an example of an irregular HyperX topology with \(L=2\), \(S_1=2\), \(S_2=4\), and \(T=3\). In this case, there are two switches in the first dimension and four switches in the second dimension, forming an irregular HyperX structure.

We note that the ToR switches connecting to QPUs in all three topologies shown in Figs.~\ref{fig:network-app}(a)-(c) are NIR switches.



\subsection{Server-centric topology}
\label{app:dcell}

\begin{figure*}
    \centering
    \includegraphics[scale=1.1]{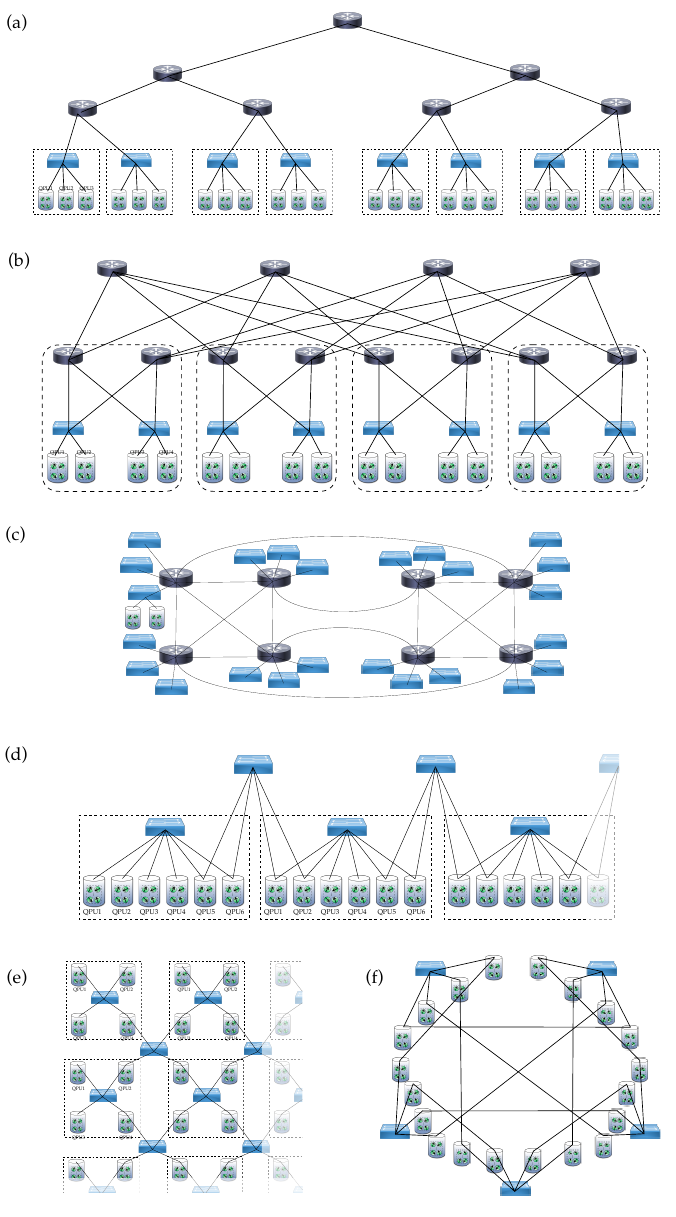}
    \caption{(a)-(c) Switch-centric network topologies: (a) Simple tree, (b) Fat-tree, and (c) HyperX. (d)-(f) Server-centric network topologies: (d) linear and (e) 2d networks introduced in this paper, and (f) DCell.}
    \label{fig:network-app}
\end{figure*}

A simplified yet more practical (for quantum network purposes) network architecture inspired by BCube is shown in Fig.~\ref{fig:network-app}(d), where neighboring racks are connected by optical switches. Because of the linear connectivity by design, we shall call it a linear network. This topology in the backbone is nothing but a linear repeater chain and we need to generate $n-1$ ebits to connect two QPUs from two racks with distance $n$. However, this constraint can be improved by modifying the rack connectivities (but not all-to-all as in the case of BCube) such as the 2D network shown in Fig.~\ref{fig:network-app}(e) (which is kind of similar to the original work of Pant et al.~\cite{pant2019routing}). 

Another example we consider is the DCell architecture\cite{guo2008dcell}, which can scale to very large number of interconnected QPUs
using switches and QPUs with very few ports. The most basic element of a DCell, which is called DCell$_0$, consists of $n$ QPUs
and one $n$-port switch. Each QPU in a DCell$_0$ is connected to the switch in the same
DCell$_0$.
The first step is to construct a DCell$_1$ from
several DCell$_0$s. Each DCell$_1$ has $n+1$ DCell$_0$s, and each QPU of every DCell$_0$
 is connected to a QPU in another DCell$_1$. As a result, the
DCell$_0$s are connected to each other, with exactly one link between every pair of
DCell$_0$s. A similar procedure is used to construct a DCell$_k$ from several DCell$_{k-1}$s.
In a DCell$_k$, each QPU will eventually have $k+1$ links: the first link or the level-$0$
link connected to a switch when forming a DCell$_0$, and level-$i$ link connected to
a QPU in the same DCell$_i$ but a different DCell$_{i-1}$. Figure~\ref{fig:network-app}(f) shows
a DCell$_1$ when $n = 4$. It can be seen that the number of QPUs in a DCell grows
double-exponentially, and the total number of levels in a DCell is limited by the
number of ports on the QPUs. For example, when $n = 6$, $k= 3$, a
fully constructed DCell can comprise more than three million QPUs.

As mentioned in the main text, the routing and scheduling of entanglement in server-centric topologies remains an open problem since it involves entanglement swapping in the intermediate QPUs.

\section{Rack assignment as ILP}
\label{app:rack assignment}

In this appendix, we provide some details on how to formulate the QPU rack assignment problem as an integer linear program. This problem needs to be addressed after the circuit partitioning step in the compiler, where we still have a freedom of placing QPUs at different racks. Considering the hybrid architecture, we wish to minimize the number of inter-rack communications.

We are to assign $N$ number of QPUs to $R$ number of racks such that the weighted number of remote gates (as defined below) are minimized. Our choice of objective is inspired by the estimated total circuit fidelity given by  $F_\text{est.} = \prod_i F_i^{n_i}$ considering we have $M$ different types of gates labeled by $i=1,\cdots, M$ and $n_i$ is the number of such gates. Upon taking the logarithm of this quantity we get a linear utility function 
\begin{align}
    U_\text{Fid} = \sum_{i=1}^M n_i \log F_i.
\end{align}
It is customary to consider one gate type as baseline (call it $i=1$) and define an objective function as a weighted sum as follows
\begin{align}
    C_\text{Fid} = \sum_{i=1}^M n_i \left(\frac{\log F_i}{\log F_1}\right),
\end{align}
which does not depend on the log base. For instance, for hybrid architectures we have two types of remote gates, intra-rack and inter-rack,
\begin{align}
    \label{eq:app-logfid-hybrid}
    C_\text{Fid}^{\text{(hyb)}} = n_\text{loc} + n_\text{intra} \left(\frac{\log F_\text{intra}}{\log F_\text{loc}}\right) + n_\text{inter} \left(\frac{\log F_\text{inter}}{\log F_\text{loc}}\right).
\end{align}
We note that the original utility function $U_\text{Fid}$ is to be maximized, while the cost function $C_\text{Fid}$ derived by dividing $U_\text{Fid}$ by $\log F_1$ is to be minimized because $\log F_1 \leq 0$.
In addition, since gate fidelities are close to one, they are often 
described by infidelities instead, $\epsilon_i = 1 -F_i \ll 1$. In this case, we can approximate the log and define the objective function as
\begin{align}
    \tilde{C}_\text{Fid} = \sum_{i=1}^M n_i \left(\frac{\epsilon_i}{\epsilon_1}\right).
\end{align}
Here, we need to minimize the number of inter-rack communications by minimizing the objective function in Eq.~(\ref{eq:app-logfid-hybrid}), where we shall drop $n_\text{loc}$ term since it is a constant. To this end, we formulate the following integer linear programming problem 
\begin{align} 
\min_{x_{ir}}  & \left(\frac{\log F_\text{inter}}{\log F_\text{loc}}\right)\sum_{ij} e_{ij} \left(1-\frac{1}{2}\sum_r y_{ijr} \right) \nonumber \\
&+  \left(\frac{\log F_\text{inter}}{\log F_\text{loc}}\right) \sum_{ijr} \frac{1}{2} e_{ij} y_{ijr} 
\label{problem:rack}
\\
\text{s.t.}& 
\nonumber\\
& \sum_{r=1}^{R}  x_{ir}   \leq 1, \quad \quad \quad \quad  \quad \quad \quad \quad \quad 
\label{cons:using_at_most_1_rack}
1 \leq i \leq N
\\
& \sum_{i=1}^N  x_{ir}   \leq n_\text{ToR}, \ \quad \quad  \quad \quad \quad \quad \quad \label{cons:tor}  
1 \leq r \leq R
\\
& y_{ijr} = (x_{ir} \oplus x_{jr}),  \quad \quad \quad  \quad \quad \quad \quad \quad \label{cons:xor}  
\\
\quad & x_{ir}  \in \{0,1\}, \ \ \quad \quad  \quad 
1 \leq i \leq N, \ 1 \leq r \leq R
\end{align}
where $e_{ij}$ denote the number of remote gates between $i$-th and $j$-th QPUs, and our binary decision variables are $x_{ir}$, a matrix indicating the rack assignment for QPU $i$, i.e., $x_{ir}=1$ means QPU $i$ is placed at rack $r$. In other words, the $R$-component vector $\boldsymbol{x}_i = (x_{i1},\cdots,x_{iR})^T$ is a one-hot vector which indicates the location of the $i$-th QPU. The one-hot vector condition is implemented by (\ref{cons:using_at_most_1_rack}). There are at most $n_\text{ToR}$ spots at each rack, this constraint is implemented by (\ref{cons:tor}). We keep track of inter-rack remote gates by the variable $y_{ijr}$ which is calculated as an exclusive-or of the two vectors associated with $i$-th and $j$-th QPUs; i.e., $\boldsymbol{y}_{ij}$ as a vector contains all zeros if the two QPUs belong to the same rack, otherwise contains two non-zero entries. 
Finally, we use the one norm $\norm{\boldsymbol{y}_{ij}}=\sum_r y_{ijr}$ to count the number of inter-rack and intra-rack remote gates.

\section{Resource management as ILP}
\label{app:dqc routing}

In this appendix, we present an integer linear programming formulation of DQC routing or resource management.
The general problem statement is as follows: Given a quantum circuit, find a sequence of gate executions which minimize the switching events and maximize the number of parallel ebit generations.

The latter problem in general can be formulated as a multi-commodity max flow problem which can be implemented as an integer linear programming as we explain below.
Minimizing switching events on the other hand, requires an algorithm which looks ahead (deeper) into the execution sequence which is more complex and out of the scope of the current work.

We start with introducing some notations:
Let $S=S_\text{ToR}\cup S_\text{C}$ be the set of optical switches,  which in turn can be decomposed 
into the top-of-rack $S_\text{ToR}$ and core (telecom) $S_\text{C}$ switches, respectively,
$G$ be the set of remote two-qubit gates, which can similarly be decomposed as 
$G= G_\text{ToR}\cup G_\text{Int}$ into the intra-rack (through top-of-rack) and inter-rack two-qubit gates.
Finally, $Q=\{1,\cdots,N\}$ denotes the set of QPUs. The rest of the system parameters are summarized in Table~\ref{tab:ilp-variables}.

As mentioned, we have two kinds of remote gates: inter-rack and intra-rack.
For each inter-rack remote gate $g$, we introduce a binary variable 
$x^{g}_p$ associated with a path $p$  connecting the respective QPUs. Here, $p$ also includes the BSM to be used on that path. In other words, if there are $n$ switches equipped with BSM devices on path $p$, we have $n$ number of variables  $x_{p_1}^g, x_{p_2}^g, \cdots, x_{p_n}^g$.
For each intra-rack remote gate $g$, we consider another binary variable
$y^{g}$ which unlike the inter-rack gates does not have a associated path since there is only one shortest optical path through the top-of-rack switch. We note that if $y^g=1$, it takes up one of the NIR BSMs on the corresponding ToR switch.

\begin{table}
    \centering
    \begin{tabular}{c|l}
        \hhline{==}
        Variable & Description \\
        \hline
         $C_q$ & Communication qubits  \\
         $B_s$ & Ports on the beam splitter  \\
         $M_s$ & BSM devices \\
         $E_s$ & Entanglement sources \\
         $D_s$ & Photon detectors \\    
        \hhline{==}
    \end{tabular}
    \caption{Notation used in ILP for the number of various network components. Subscripts $q$ and $s$ refer to QPUs and switches, respectively.}
    \label{tab:ilp-variables}
\end{table}

The goal of ILP is to find a solution (assign values to $x^{g}_p$ and $y^{g}$) by maximizing the number of ebits which can be generated in parallel. To this end, we define the objective function as a weighted sum over inter- and intra-rack paths where weight factors $w^g$ can be used (as input)  to implement some sort of gate priority in terms of their importance in the DAG (c.f. main text), or other factors. Hence, the ILP is given by
\begin{align} 
\max_{x^{g}_p, y^{g}}  & \sum_{g \in G_\text{Int}, p\in P^g} w^g x^{g}_p 
  + \sum_{g \in G_\text{ToR}} w^g y^{g} 
\label{problem:exhaustive_search_pathbased}
\\
\text{s.t.}& 
\nonumber\\
& \sum_{\substack{p \in P^g}
}  x^{g}_{p}   \leq 1 \quad \quad \quad \quad  \quad \quad \quad \quad \quad \forall{g\in G}
\label{cons:using_at_most_1_path}
\\
&\sum_{q \in g, p \in P^g} x^{g}_{p} 
+ \sum_{q \in g} y^g + 
\leq C_q,
\label{cons:commq}   \quad \quad \quad \forall q\in Q  
\\
& \sum_{g \in G_\text{ToR}, s \in g} y^{g}  \leq B_s/2,
\label{cons:BS-tor}  \quad \quad  \quad \quad \quad \forall s\in S_\text{ToR}  
\\
& \sum_{g \in G_\text{ToR}, s \in g} y^{g} \leq M_s,
\label{cons:BSM-tor}  \quad \quad  \quad \quad \quad \quad \forall s\in S_\text{ToR}  
\\
& \sum_{\substack{g \in G \\
p \in P^g |s \in p
}
} x^{g}_{p}  \leq E_s,
\label{cons:ES-tor} \quad \quad \quad  \quad \quad \quad \quad \forall s\in S_\text{ToR}  
\\
& \sum_{\substack{g \in G \\
p \in P^g |s \in p
}
} x^{g}_{p}  \leq D_s,
\label{cons:PD-tor} \quad \quad \quad  \quad \quad \quad \quad \forall s\in S_\text{ToR}  
\\
& \sum_{\substack{g \in G \\
p \in P^g |s \in p
}
} x^{g}_{p}  \leq M_s,
\label{cons:BSM-core}  \quad \quad  \quad \quad \quad \quad \forall s\in S_\text{C}  
\\
\quad & x^{g}_{p}  \in \{0,1\} ,   \quad \quad \quad \quad \quad  \quad \forall g \in G,  p \in P^{g}
\\
\quad & y^{g}  \in \{0,1\} ,  \quad \quad \quad \quad \quad  \quad \forall g \in G_\text{ToR}
\end{align}
We now go over the constraints. Eq.~(\ref{cons:using_at_most_1_path}) implies only one path per remote gate $g$ is required. Eq.~(\ref{cons:commq}) imposes an upper bound on the number of paths connected to a QPU $q$ participating in gate $g$ because of the limited number of communication qubits $C_q$. Eqs.~(\ref{cons:BS-tor}) and (\ref{cons:BSM-tor}) are required for intra-rack communications due to the number of ports on the beam splitters and NIR BSMs on ToR switches, respectively. To make sure there is no competition between the number of beam splitters and NIR BSMs, we can simply set $B_s = 2 M_s$ and reduce these two constraints to one. 
Similarly, Eqs.~(\ref{cons:ES-tor})-(\ref{cons:BSM-core}) are required for inter-rack communications which impose constraints due to limited number of entanglement sources, photon detectors (for scatterers) and telecom BSM devices.

In principle, we need to consider all possible paths between the QPUs which grows as $N!$ with the number of QPUs. However, enumerating all paths is not useful in practice, since majority of paths are long and not desirable. So, we can only limit ourselves to the set of shortest paths for each two-qubit gate. This is at the expense of more resource contention which leads to additional rounds of switching. Even limiting our solution into this smaller subspace of paths, the above ILP may still remain degenerate and have many solutions due to symmetries of the network graph. We note that all these solutions are all equally acceptable since we already limit our search to the shortest paths.

\bibliography{refs.bib}

\end{document}